\documentclass[12pt,a4paper]{article}
\usepackage{amsmath}
\usepackage{amsxtra}
    \usepackage{amstext}
    \usepackage{amssymb}
    \usepackage{latexsym}
    \usepackage{graphicx}
\usepackage{color}
\usepackage{graphics}

\newcommand{\hepth}[1]{{\tt hep-th/#1}}

\newcommand{\p}{\vspace{6pt}\noindent}

\advance\textheight by \topskip
\makeatletter

\@addtoreset{equation}{section}
\def\section{\@startsection {section}{1}{\z@}{-8.5ex plus -1ex minus
 -.2ex}{3.3ex plus .2ex}{\large\bf}}
\def\subsection{\@startsection{subsection}{2}{\z@}{-3.25ex plus
 -1ex minus -.2ex}{1.5ex plus .2ex}{\bf}}
\def\subsubsection{\@startsection{subsubsection}{3}{\z@}{-3.25ex plus%
 -1ex minus -.2ex}{1.5ex plus .2ex}{\sl}}

\begin{document}

\begin{titlepage}
\vspace*{-2cm}
\begin{flushright}
\end{flushright}

\vspace{0.3cm}

\begin{center}
{\Large {\bf }} \vspace{1cm} {\Large {\bf Integrable defects in affine Toda field theory and infinite dimensional representations of quantum groups}}\\
\vspace{1cm} {\large  E.\ Corrigan\footnote{\noindent E-mail: {\tt
edward.corrigan@durham.ac.uk}} and
C.\ Zambon\footnote{\noindent E-mail: {\tt cristina.zambon@durham.ac.uk}} \\
\vspace{0.3cm}
{\em Department of Mathematical Sciences \\ University of Durham, Durham DH1 3LE, U.K.}} \\

\vspace{2cm} {\bf{ABSTRACT}}\\

\vspace{.5cm}

\end{center}
\p Transmission matrices for two types of integrable defect are calculated explicitly, first by solving directly the nonlinear transmission Yang-Baxter equations, and second by solving a linear intertwining relation between a finite dimensional representation of the relevant Borel subalgebra of the quantum group underpinning the integrable quantum field theory  and a particular infinite dimensional representation expressed in terms of sets of generalized `quantum' annihilation and creation operators.  The principal examples analysed are based on the $a_2^{(2)}$ and $a_n^{(1)}$ affine Toda models but examples of similar infinite dimensional representations for quantum Borel algebras for all other affine Toda theories are also provided.\\

\vfill
\end{titlepage}

\section{Introduction}

\p This article is a further step in the program  to provide classical and quantum descriptions of integrable defects that can be supported within the massive Toda field theories. The insistence on maintaining integrability leads to strong constraints on the types of defect that can be sustained. Many years ago, Delfino, Mussardo and Simonetti demonstrated within a quantum field theory possessing non-trivial bulk scattering that compatibility with the bulk S-matrix forces a defect to be purely transmitting \cite{Delf94}. Later, a classical account of a purely transmitting defect based on a Lagrangian description was provided in \cite{bczlandau}. There a defect was introduced as a field discontinuity together with a set of sewing conditions relating fields across the discontinuity. The first example of a defect of this type within a quantum field theory was studied in the sine-Gordon model \cite{konik97, bczsg05,Habibullin2007, Nemes2009}, see also \cite{kundu}. In \cite{cz09}, it was shown that this kind of defect, called a type-I defect, was allowed classically only within the Toda models based on the $a_n^{(1)}$ root data. Some aspects of the classical and quantum descriptions of defects within these models have  been provided \cite{bcz2004, cz2007}.

\p The strong limitation on the variety of Toda models able to support type-I defects motivated the search for an alternative classical description of a purely transmitting defect. In \cite{cz2009} the type-II defect, which added an additional degree of freedom at the location of the defect,
was proposed. This new setting proved to be suitable not only  for describing a purely transmitting defect within the $a^{(2)}_2$ Toda model (variously described as the Tzitz\'eica, or Bullough-Dodd, or Mikhailov-Zhiber-Shabat model), which had been excluded previously, but it also provided additional types of defect for the sine-Gordon and other Toda models. A quantum description of a type-II defect within the sine-Gordon model, effectively generalising the transmission matrix for the type I defect \cite{konik97,bczsg05}, was suggested in \cite{cz2010}.

\p The transmission matrices that describe the scattering of a soliton with type-I or  type-II defects within affine Toda models are infinite dimensional, reflecting the fact that  defects are able to store topological charge. From an algebraic point of view, the transmission matrices are related to infinite dimensional representations  of the Borel subalgebra of the quantum groups underpinning the models themselves. For the sine-Gordon model, the precise relationship between the transmission matrices for the type-I or type-II defects and certain infinite dimensional representations of the algebra generated by a pair of creation and annihilation operators has been provided by Weston \cite{weston} (see also \cite{Habibullin2007}). The basic tool is the intertwining condition between an infinite dimensional representation and a spin $1/2$ two-dimensional representation of the algebra $U_q(a_1^{(1)})$. This relation is linear and therefore generally easier to handle than the quadratic transmission Yang-Baxter equation proposed in \cite{Delf94}.

\p Representation theory turns out to be a a powerful tool to investigate suitable transmission matrices for Toda field theories with defects. In this article, the transmission matrix describing the scattering of an $a^{(2)}_2$ Toda soliton with a type-II defect is constructed using both methods. First,  the quadratic `triangle relations' are solved explicitly and, second,  the linear intertwining equations are solved following the construction of a suitable infinite dimensional representation of the Borel subalgebra of  $U_q(a_2^{(2)})$ in terms of a pair of generalized creation and annihilation operators. The results are compared and linked to the classical description of the problem provided previously in \cite{cz2009}. An interpretation of the additional degrees of freedom appearing in the Lagrangian description is suggested in terms of the representation theory.

\p The method is also applied to the $a^{(1)}_n$ Toda models. Suitable infinite dimensional representations of the relevant quantum algebras are constructed and the solutions of the linear intertwining relations are obtained. Interestingly, the representation that is most appropriate for the defect scenario, in the sense that the cyclic symmetry of the Dynkin-Kac diagram is preserved, is not precisely the same as those that have been developed previously in other contexts, for example \cite{Hayashi, bazhanov2002}. The results of these calculations are then compared with results presented previously for type-I defects within the $a^{(1)}_n$ Toda models, obtained by solving the transmission Yang-Baxter equation directly. In addition,  suitable transmission matrices for type-II defects are provided. In fact, the classical Lagrangian description of  type-II defects within the $a^{(1)}_n$ Toda models is also provided here for the first time.

\p Finally, a universal rule for building suitable infinite dimensional representations of the Borel subalgebras appropriate to any of the affine Toda models  is proposed in an appendix. These constructions, given in terms of sets of pairs of generalized `quantum' creation and annihilation operators, are an interesting diversion from the main topic of the paper and explicit examples are provided for all models. It is also interesting to explore which finite dimensional representations are achievable (since the transmission matrices can contain - though not always - finite dimensional S-matrices). This aspect was noted in \cite{cz2010} in the context of the sine-Gordon model  and some further comments are made in section (6) in the context of the $a_n^{(1)}$ affine Toda models. More complete  detail on these constructions and their properties will be presented elsewhere.

\section{The classical type-II defect}
\label{classicalsection}

\p In this section a few facts concerning the type-II defect will be recalled. The original ideas and a number of results can be found in \cite{cz2009}, though the Lagrangian starting point used in the present article is slightly different. As explained in \cite{cz2010}, the two settings are completely equivalent and the Lagrangian density describing two scalar fields $u$ and $v$ defined in the two regions $x<0$ and $x>0$, respectively, and separated by a defect at $x=0$, may be taken to be
\begin{equation}\label{defectlagrangian}
{\cal L}=\theta(-x)\,{\cal L}_u + \theta(x)\,{\cal L}_v +\delta(x)\left(2 q\cdot \lambda_t -{\cal D}(\lambda,u,v)\right),
\end{equation}
where the quantity $\lambda(t)$ is confined to $x=0$. The defect potential ${\cal D}$ satisfies the following
\begin{equation}\label{Dconstrain}
 {\cal D}=f(p+\lambda,q)+g(p-\lambda,q),\quad \nabla_\lambda f \cdot \nabla_q g-\nabla_\lambda g \cdot \nabla_q f=U(u)-V(v),
\end{equation}
with
$$ p=\frac{u+v}{2},\quad q=\frac{u-v}{2}.$$
In the latter, the fields $u(0,t)$ and $v(0,t)$ are defined by taking limits of $u(x,t)$ and $v(x,t)$ as $x\rightarrow 0$ in their respective domains. Typically, $u,\ v$ and $\lambda$ will be multi-component fields and the constraints \eqref{Dconstrain} are powerful because of the absence of $\lambda$ on the right hand side of the second relation. For the affine Toda model related to the  extended Lie algebra $a_2^{(2)}$ the fields $u$ and $v$ are single component scalar fields with bulk potentials $U$ and $V$,
$$U(u)=-\frac{m^2}{\beta^2}\,(e^{i\,\beta\,u\sqrt{2}}+2\,e^{-i\,\beta\,u/\sqrt{2}}-3),\quad
V(v)=-\frac{m^2}{\beta^2}\,(e^{i\,\beta\,v\sqrt{2}}+2\,e^{-i\,\beta\,v/\sqrt{2}}-3),$$
where $\beta$ is a real coupling constant.
The defect potential satisfying \eqref{Dconstrain} is given by
\begin{eqnarray}\label{defectpotentiala22}
{\cal D}(\lambda,p,q)&=&\frac{\sqrt{2}\,m\,\sigma}{\beta^2} \, \left(e^{i(p+\lambda)\,\beta/\sqrt{2}}+e^{-i(p+\lambda)\,\beta/2\sqrt{2}}
\,\left(e^{iq\,\beta/\sqrt{2}}+e^{-iq\,\beta/\sqrt{2}}\right)\right)\nonumber \\
&&+\frac{\sqrt{2}\,m}{2\,\sigma\,\beta^2} \, \left(8\,e^{-i(p-\lambda)\,\beta/2\sqrt{2}}+e^{i(p-\lambda)\,\beta/\sqrt{2}}
\,\left(e^{iq\,\beta/\sqrt{2}}+e^{-iq\,\beta/\sqrt{2}}\right)^2\right),
\end{eqnarray}
where $\sigma$ is the defect parameter.

\p
 The delay experienced by a soliton travelling through a defect was calculated in \cite{cz2009} and its precise form depends on the initial conditions of the fields $u$, $v$ and $\lambda$. It is worth remembering that the Tzitz\'eica model has complex soliton solutions distinguished by two different topological charges and these will be referred to as the classical soliton and anti-soliton. The delay of most interest for the present investigation is $z$, given by:
\begin{equation}\label{classicaldelaya22}
z=\coth\left(\frac{\theta-\eta}{2}-\frac{i\pi}{12}\right)
\coth\left(\frac{\theta-\eta}{2}+\frac{i\pi}{12}\right),\qquad \sigma=\sqrt{2}\,e^{-\eta}.
\end{equation}
Note that if $\eta$ is real (and the rapidity $\theta$ is always assumed to be real), then the
delay \eqref{classicaldelaya22} is real and positive. This means a soliton passing
the defect cannot convert to an antisoliton (or vice versa), or be trapped by the defect. On the other hand, the parameter $\eta$ might be complex and then, for certain choices of $\eta$, a soliton may convert to an antisoliton or be absorbed by the defect, illustrating behaviour already exhibited by the basic type-I defect in the sine-Gordon model.

\p  The type-II defect was extensively investigated for the sine-Gordon model in \cite{cz2010}, both in the classical and quantum contexts. Here, its generalisation for all the remaining affine Toda models in the $a_n^{(1)}$ series will be presented for the first time. In these cases, the fields $u$, $v$  and $\lambda$ are multi-component scalar fields. The bulk potentials for the fields $u$ and $v$ are:
\begin{equation}
U(u)=-\frac{m^2}{\beta^2}\sum_{j=0}^{n}\left(e^{i\beta\alpha_j\cdot u}-1\right),\quad
V(v)=-\frac{m^2}{\beta^2}\sum_{j=0}^{n}\left(e^{i\beta\alpha_j\cdot v}-1\right),
\end{equation}
where $\alpha_j,\ j=1,\dots n$, are a set of simple roots and $\alpha_0=-\sum_{j=1}^n \alpha_j$.

\p The appropriate Lagrangian density is given by \eqref{defectlagrangian} together with two slightly different possibilities for the defect potential $\mathcal{D}$. They will be referred to as setting A and setting B. In setting A the defect potential is:
\begin{equation}\label{defectpotentiala21}
\mathcal{D}(\lambda,p,q)=\frac{m}{\beta^2}\sum^{n}_{j=0}\left(\sigma e^{i\beta\alpha_j\cdot(p+\lambda)/2}A_j(q)+
\frac{1}{\sigma}e^{i\beta\alpha_j\cdot(p-\lambda)/2}A_{j+1}(q)\right),
\end{equation}
where
$$ A_j(q)=\gamma \, e^{i\beta\alpha_j\cdot G q/2}+\frac{1}{\gamma}\,e^{-i\beta\alpha_j\cdot G q/2},
\qquad p=\frac{u+v}{2},\quad q=\frac{u-v}{2}.$$
The constant matrix $G$ is constructed as follows,
$$G=2\sum_{a=1}^n (w_a-w_{a+1})\,w_a^T,\quad \alpha_i\cdot w_j=\delta_{ij}\quad i,j=1,\dots,r$$
where $w_i$ are the fundamental weights of the Lie algebra $a_n^{(1)}$, and $\sigma$, $\gamma$ are the two defect parameters. In setting B the defect potential is given by the expression \eqref{defectpotentiala21} with $p$ replaced by $-p$ and the matrix $G$ replaced by the matrix $-G$.
In \cite{bcz2004} the classical type I-defect for the Toda models in the $a_n^{(1)}$ series were presented. There, two different Lagrangian settings for the defect were also introduced. The two settings A and B for the type-II defect can be seen as stemming from those two possibilities. It is worth recalling that the existence of two different settings has interesting consequences for the conservation of the classical charges, as noted in \cite{bcz2004}.

\p As was argued in \cite{cz2010} for the sine-Gordon model,  the type-II defect may be interpreted as two type-I defects \lq fused' together at the same point. However, unlike the sine-Gordon model, the Dynkin-Kac diagrams for the affine Toda models in the $a_n^{(1)}$ series have a cyclic symmetry that can be seen as a `clockwise' or an `anticlockwise' permutation of the roots. Then, to each of them can be associated a slightly different type-I defect (see \cite{bcz2004,cz2007} for further details on this point) and it is necessary to have one of each to achieve the `fusion' process mentioned previously.

\p Finally, there are generalizations of these ideas to some of the other affine Toda field theories though not yet to all of them. A discussion of these generalizations would be a digression from the main topic of the paper and is deferred to another occasion.

\section{The $a_2^{(2)}$ transmission matrix: the quadratic equation}
\label{thequadraticequation}

\p In this section, the transmission matrix $T$ describing the scattering between a soliton and a type-II defect will be calculated. The procedure adopted relies on solving the transmission-Yang-Baxter equation \cite{Delf94}, or \lq triangle relations', namely
\begin{equation}\label{STT}
S_{a\,b}^{mn}(\Theta)\,T{_{n\alpha}^{t\beta}}(\theta_1)\,
T{_{m\beta}^{s\gamma}}(\theta_2)=T{_{b\,\alpha}^{n\beta}}(\theta_2)\,
T{_{a\,\beta}^{m\gamma}}(\theta_1)\,S_{mn}^{st}(\Theta),\qquad \Theta=\theta_1-\theta_2,
\end{equation}
where $S$ is the bulk scattering matrix for the $a_2^{(2)}$ affine Toda model. Effectively, once the the $S$-matrix is given these are a set of quadratic recurrence relations for the components of the infinite-dimensional matrices $T$. In the quantum version of the $a_2^{(2)}$ affine Toda model the fundamental particle is represented by a three-component soliton, unlike the classical situation where only two components appear as classical solutions, and the $S$-matrix describes the scattering of these solitons. A suitable $S$-matrix was proposed some time ago by Smirnov \cite{smirnov} using the Izergin-Korepin $R$ matrix \cite{IKRmatrix}.  The $S$-matrix acts on a two particle asymptotic state as follows:
$$
_{\mbox{\small{in}}}<\theta_1,a;\theta_2,b|\,S_{ab}^{dc}(\Theta)=
\,_{\mbox{\small{out}}}<\theta_2,d;\theta_1,c|,
$$
where the Roman labels take the values $+1$, $0$, $-1$ or simply ($+$, $0$, $-$) to indicate the three
fundamental solitons.
The $S$ matrix is defined in detail by,
\begin{equation}\label{a22Smatrix}
S(\Theta)=\rho_S(\Theta)P\,R(x,q),\quad x=\frac{x_1}{x_2},\quad x_i=q^{2\pi\theta_i/\xi},
 \quad \xi=\frac{2}{3}\,\left(\frac{\pi \beta^2}{8\pi-\beta^2}\right),
\end{equation}
where $\beta$ is the coupling constant, $P$ is the permutation operator and $\rho_S$ is an overall scalar factor. The latter can be found in \cite{smirnov} but will not be needed here. The matrix $R$ appearing in \eqref{a22Smatrix} is invariant under the action of the $U_q (a_2^{(2)})$ algebra, meaning it intertwines between two representations of the algebra, as follows,
$$R(x_1/x_2,q):V_{x_1}\otimes V_{x_2}\rightarrow V_{x_2}\otimes V_{x_1}.$$
In detail, it is given by,
$$R=(x^{-1}-1)\,q^{3}\,R_{12}+(1-x)\,q^{-3}\,R_{21}^{-1}+
q^{-5}\,(q^{4}-1)\,(q^{6}+1)\,P,\quad R_{21}=P\,R_{12}\,P,$$
where $R_{12}$ is the constant solution of the Yang-Baxter equations for the $U_q({sl_2})$ spin 1 representation.
Explicitly, the $R$ matrix in \eqref{a22Smatrix} can be written as follows
\begin{equation}\label{a22Rmatrix}
R=\left(
  \begin{array}{ccccccccccc}
    c & 0 & 0 &|& 0 & 0 & 0 &|& 0 & 0 & 0 \\
    0 & b & 0 &|& \tilde{e} & 0 & 0&| & 0 & 0 & 0  \\
     0 & 0 & d &|& 0 & \tilde{g} & 0&| & \tilde{f} & 0 & 0  \\
    - & - & - &|& - & - & -&| & - & - & -  \\
    0 & e & 0&| & b & 0 & 0 &|& 0 & 0 & 0  \\
    0 & 0 & g &|& 0 & a & 0 &|& \tilde{g} & 0 & 0  \\
     0 & 0 & 0 &|& 0 & 0 & b&| & 0 & \tilde{e} & 0  \\
      - & - & - &|& - & - & -&| & - & - & -  \\
     0 & 0 & f &|& 0 & g & 0 &|& d & 0 & 0  \\
     0 & 0 & 0 &|& 0 & 0 & e &|& 0 & b & 0  \\
     0 & 0 & 0 &|& 0 & 0 & 0 &|& 0 & 0 & c  \\
  \end{array}
\right),
\end{equation}
with
\begin{eqnarray}
b\phantom{_+}&=&b_++b_-,\quad a=b_++b_-+h,\quad d=q^{2}\,b_+ +q^{-2}\,b_-,\quad c=q^{-2}\,b_+ +q^{2}\,b_- +h,\nonumber \\
\tilde{g}\phantom{_+}&=&\left(x^{-1}-1\right)\,(q^{4}-1),\quad \tilde{e}=q\,\tilde{g}+h,\quad
\tilde{f}=\tilde{g}\,(q-q^{-1})+h,\nonumber \\
g\phantom{_+}&=&(x-1)\,(1-q^{-4}),\quad e=q^{-1}\,g+h,\quad
f=g\,(q^{-1}-q)+h,\nonumber \\
b_+&=&q^{-3}\,(1-x),\quad b_-=q^{3}\,\left(x^{-1}-1\right),\quad h=-(q-q^{-1})+(q^5-q^{-5}).
\end{eqnarray}

\p The transmission matrix has two different kinds of labels. The Roman labels indicate solitons, as before, while the Greek labels are integers referring to the defect. The equation \eqref{STT} has been solved explicitly and directly by assuming that the total topological charge of the system (that is, soliton plus defect) is conserved, meaning a soliton scattering with a defect is permitted to change its topological charge. The general solution is:
\begin{equation}\label{Ta22solution}
T_{a\alpha}^{b\beta}=\rho(\theta)\,\left(
  \begin{array}{ccc}
    (\varepsilon^2\,q^{2\,\alpha}+\tau^2 \,q^{-2\,\alpha}\,x)\,\delta^{\beta}_{\alpha}& \varepsilon\mu(\alpha) \,\delta^{\beta-1}_{\alpha} & M(\alpha)\,\delta^{\beta-2}_{\alpha} \\
    \tau\lambda(\alpha)\,x\,\delta^{\beta+1}_{\alpha}& (\tilde{\tau}\,\varepsilon+\tau\,\tilde{\varepsilon}\,x)\,\delta^{\beta}_{\alpha} & \tilde{\tau}\,\,\mu(\alpha) q^{-2\,\alpha-1} \,\delta^{\beta-1}_{\alpha} \\
    L(\alpha)\,x\,\delta^{\beta+2}_{\alpha}& \tilde{\varepsilon}\,q^{2\,\alpha-1}\,\lambda(\alpha)\,x \,\delta^{\beta+1}_{\alpha} & (\tilde{\varepsilon}^2\,q^{2\,\alpha}\,x+\tilde{\tau}^2 \,q^{-2\,\alpha})\,\delta^{\beta}_{\alpha} \\
  \end{array}
\right),
\end{equation}
where
$$M(\alpha)=\mu(\alpha)\,\mu(\alpha+1)\,
    \frac{q^{-2\,\alpha-1}}{1+q^{2}}, \quad L(\alpha)= \lambda(\alpha)\,\lambda(\alpha-1)\,\frac{q^{2\,\alpha-1}}{1+q^2},$$
and
\begin{equation}\label{mulambdarelation}
\mu(\alpha)\,\lambda(\alpha+1)=(q+q^{-1})\,\left(\tau\,\tilde{\tau}\,q^{-2\,\alpha-1}
+\varepsilon\,\tilde{\varepsilon} \,q^{2\,\alpha+1}\right),\qquad q=e^{i 4\pi^2/\beta^2}.
\end{equation}
Clearly this solution has plenty of freedom, represented by four free parameters and one of the functions $\lambda(\alpha)$ or $\mu(\alpha)$. Note, a similarity transformation will be unable in general to eliminate such an arbitrary function. Part of the challenge is to reduce the number of free parameters and then identify the remainder with parameters appearing in the Lagrangian model. Note also, at this stage, the overall scalar factor $\rho$ is undetermined.

\p As usual, to select a suitable transmission matrix, additional constraints must be imposed and they are provided by a crossing condition and by examining the pole structure of the $S$-matrix, as explained below. Note that the model is not unitary in a `quantum mechanics sense'. First, the crossing condition reads
\begin{equation}\label{crossing}
T^{b\,\beta}_{a\,\alpha}(\theta)=\tilde{T}^{\bar a\,\beta}_{\bar b\,\alpha}(i\pi-\theta),
\end{equation}
where the matrix $\tilde{T}$ is defined as follows
$$T^{b\,\beta}_{a\,\alpha}(\theta)\tilde{T}^{c\,\gamma}_{b\,\beta}(-\theta)
=\delta^{c}_{a}\delta^{\gamma}_{\alpha},$$
that is
\begin{eqnarray*}
&&\tilde{T}_{a\alpha}^{b\beta}(-\theta)=\tilde{\rho}(-\theta)\\
&&\times\,\left(
  \begin{array}{ccc}
    (\tilde{\tau}^2\,q^{-2\,\alpha-4}-\tilde{\varepsilon}^2 \,q^{2\,\alpha+2}\,x)\,\delta^{\beta}_{\alpha}& -\tilde{\tau} \,\mu(\alpha)\,q^{-2\alpha-4}\,\delta^{\beta-1}_{\alpha} & M(\alpha) q^{-2}\,
    \delta^{\beta-2}_{\alpha} \\
    \tilde{\varepsilon}\,\lambda(\alpha)\,x\,q^{2\alpha}\,
    \delta^{\beta+1}_{\alpha}& (\tilde{\tau}\,\varepsilon\,q^{-4}-\tau\,\tilde{\varepsilon}\,
    q^2\,x)\,\delta^{\beta}_{\alpha} & -\varepsilon\, q^{-3}\,\mu(\alpha) \,\delta^{\beta-1}_{\alpha} \\
    -L(\alpha)\,x\,\delta^{\beta+2}_{\alpha}& \tau\,q\,\lambda(\alpha)\,x \,\delta^{\beta+1}_{\alpha} & \left(\varepsilon^2\,q^{2\,\alpha-4}-\tau^2 \,x\,q^{-2\,\alpha+2}\right)\,\delta^{\beta}_{\alpha} \\
  \end{array}
\right),
\end{eqnarray*}
with
$$\tilde{\rho}(-\theta)=(\rho(\theta-i\pi)\,(\tilde{\tau}
\varepsilon+\tau\tilde{\varepsilon}q^{-4}x)\,
(\tilde{\tau}\varepsilon-\tau\tilde{\varepsilon}q^{-2}x))^{-1}.$$
Then, the crossing condition \eqref{crossing} is satisfied, at least up to a similarity transformation on the matrix part of $T$, provided the following constraint on the overall factor $\rho$ holds
\begin{equation}\label{rhocons1}
\rho(\theta)\,\rho(\theta-i\pi)\,(\tilde{\tau}\varepsilon+\tau\tilde{\varepsilon}q^{-4}x)
\,(\tilde{\tau}\varepsilon-\tau\tilde{\varepsilon}q^{-2}x)=1.
\end{equation}
Second, the $S$ matrix has simple poles at $\Theta=2 i\pi/3-i\xi m$ where $m$ is zero or a positive integer \cite{smirnov}.
These poles correspond to three particles, which belong to a spin one representation of $U_q(sl_2)$. For $m=0$ the pole corresponds to the creation of a fundamental soliton as a bound state of two other fundamental solitons. Hence, the bootstrap calculated at this point can be use to obtain consistency conditions for the $T$ matrix, and hence additional constraints on the parameters.
The bootstrap reads
\begin{equation}\label{bootequ}
c_{ab}^f\,T_{f\alpha}^{c\beta}(\theta)=\sum_{d,c}\,T_{b\alpha}^{d\gamma}(\theta_b)\,
T_{a\gamma}^{e\beta}(\theta_a)\,c_{ed}^c,
\end{equation}
with
$$\theta_a=\theta+\frac{\pi i}{3},\quad \theta_b=\theta-\frac{\pi i}{3}.$$
Some of the coupling ratios can be determined from the $S$ matrix \eqref{a22Smatrix} and they are:
$$\frac{c_{0+}^+}{c_{+0}^+}=\frac{c_{-0}^-}{c_{0-}^-}=-q^2,
\quad \frac{c_{00}^0}{c_{+-}^0}=-\frac{c_{00}^0}{c_{-+}^0}=-\frac{(q^2-1)}{q},\quad \frac{c_{+-}^0}{c_{-+}^0}=-1.$$
The bootstrap condition provides further coupling ratios and the following constraint on the overall
function $\rho$ of the $T$ matrix
\begin{equation}\label{rhocons2}
\rho(\theta)=(\tilde{\tau}\varepsilon+\tau\tilde{\varepsilon}x)\,
\rho(\theta+i\pi/3)\,\rho(\theta-i\pi/3).
\end{equation}

\p Before solving the constraints \eqref{rhocons1} and \eqref{rhocons2}, it is desirable to reduce the number of free constants appearing in the $T$ matrix \eqref{Ta22solution}. The results already obtained suggest setting
$$\varepsilon=\tilde{\tau}=1,\quad \tilde{\varepsilon}=\tau\equiv e^{-\pi\nu/\xi},$$
where $\nu$ is the one remaining free parameter. The undetermined function of $\alpha$ turns out to be unconstrained and it is tempting to relate it to the presence of the field $\lambda$ in the classical Lagrangian density \eqref{defectlagrangian}, which is only defined up to a `gauge' transformation, as was explained in \cite{cz2009, cz2010}. More comments on this issue will be provided later on.

\p Then, after redefining the function $\lambda(\alpha)$ as $\tau^2\,\lambda(\alpha)$ the $T$ matrix \eqref{Ta22solution} becomes
\begin{equation*}
T_{a\alpha}^{b\beta}=\rho(\theta)\,\left(
  \begin{array}{ccc}
    (q^{2\,\alpha}+q^{-2\,\alpha}\,\tau^2\,x)\,\delta^{\beta}_{\alpha}& \frac{F(\alpha+1)}{\lambda(\alpha+1)} \,\delta^{\beta-1}_{\alpha} & \frac{F(\alpha+1)\,F(\alpha+2)}{\lambda(\alpha+1)
    \lambda(\alpha+2)\,(1+q^{2})}
    \,q^{-2\,\alpha-1}\,\delta^{\beta-2}_{\alpha} \\
    \lambda(\alpha)\,\tau^2\,x\,\delta^{\beta+1}_{\alpha}& (1+\tau^2\,x)\,\delta^{\beta}_{\alpha} & \frac{F(\alpha+1)}{\lambda(\alpha+1)}\,q^{-2\,
    \alpha-1}\,\delta^{\beta-1}_{\alpha} \\
    \frac{\lambda(\alpha)\,\lambda(\alpha-1)}{(1+q^{2})}
    \,\tau^2\,x\,q^{2\,\alpha-1}\,\delta^{\beta+2}_{\alpha}& \lambda(\alpha)\,q^{2\,\alpha-1}\,\tau^2\,x \,\delta^{\beta+1}_{\alpha} & (q^{-2\,\alpha}+\tau^2\,x\,q^{2\,\alpha})\,\delta^{\beta}_{\alpha} \\
  \end{array}
\right),
\end{equation*}
with
\begin{equation}\label{mulambdarelation}
F(\alpha)=\left(q+q^{-1}\right)\,\left(q^{2\alpha-1}+q^{-2\alpha+1}\right).
\end{equation}
By making use of this setting, a solution to the constraints \eqref{rhocons1} and \eqref{rhocons2}
is represented by the following expression
\begin{equation}
\rho(\theta)=\hat{\rho}(\theta)\,\frac{e^{-\pi(\theta-\nu)/\xi}}{2\pi},
\end{equation}
with
\begin{eqnarray*}
\hat{\rho}(\theta)&=&\Gamma(1/2-z+\pi/3\xi)\,\Gamma(1/2-z+2\pi/3\xi)\\
&&\times \prod_{k=1}^{\infty}\,
\frac{\Gamma(1/2+z+2k\pi/\xi-4\pi/3\xi)\,\Gamma(1/2+z+2k\pi/\xi-5\pi/3\xi)}
{\Gamma(1/2+z+2k\pi/\xi-2\pi/3\xi)\,\Gamma(1/2+z+2k\pi/\xi-\pi/3\xi)}\\
&&\times\prod_{k=1}^{\infty}\,\frac{\Gamma(1/2-z+2k\pi/\xi+2\pi/3\xi)\,
\Gamma(1/2-z+2k\pi/\xi+\pi/3\xi)}
{\Gamma(1/2-z+2k\pi/\xi-2\pi/3\xi)\,\Gamma(1/2-z+2k\pi/\xi-\pi/3\xi)},\quad z=\frac{i(\theta-\nu)}{\xi}.
\end{eqnarray*}

\p  The overall factor has two interesting simple poles at
$$\theta_1=\nu-i\left(\xi+\frac{\pi}{3}\right),\quad \theta_2=\nu-i\left(\xi+\frac{2\pi}{3}\right),$$
which occur in the two gamma functions outside the infinite product. In the classical limit $\xi\rightarrow 0$ they can be compared with the poles appearing in the classical delay $z$ \eqref{classicaldelaya22}.
Then,
\begin{equation}\label{classandquanparameter}
\nu=\eta+\frac{i\pi}{2},
\end{equation}
where $\eta$ is related to the classical defect parameter.
The energies associated with these two poles can be calculated and they are:
\begin{eqnarray*} E_1&=&M\cosh\eta\cos\left(\frac{\pi}{6}-\xi\right)+i\,
M\,\sinh\eta\sin\left(\frac{\pi}{6}-\xi\right),\\
E_2&=&M\cosh\eta\cos\left(\frac{\pi}{6}+\xi\right)-i M\sinh\eta\,\sin\left(\frac{\pi}{6}+\xi\right),
\end{eqnarray*}
where $M$ is the soliton mass.
These poles can be interpreted as unstable bound states in the quantum theory provided the imaginary parts of their energies are negative. This happens, for instance, if $\pi/6<\xi<2\pi/3$ for the bound state described by the pole $\theta_1$ and if $-\pi/6<\xi<\pi/3$ for the pole $\theta_2$.

\p Finally, the $S$ matrix has also poles at $\Theta=i\pi-i\xi m$ with $m$ a positive integer and these correspond to breather bound states. The transmission factor for the lightest breather may be compared with the linearized classical defect problem since the lightest breather corresponds to the quantum particle described by the fundamental bulk scalar field. This transmission factor can be calculated making use of the bootstrap equation \eqref{bootequ} with
$$\theta_a=\theta+\frac{i}{2}\,(\pi-\xi),\quad \theta_b=\theta-\frac{i}{2}\,(\pi-\xi).$$
The coupling ratios can be determined from the $S$ matrix and are:
$$\frac{c_{+-}^{1}}{c_{00}^1}=\frac{c_{00}^1}{c_{-+}^1}=-\frac{1}{q},
\quad \frac{c_{+-}^1}{c_{-+}^1}=\frac{1}{q^2}.$$
After some calculation, the transmission factor for the lightest breather is
\begin{equation}
{^1}T=\tanh\left(\frac{(\theta-\nu)}{2}-\frac{i\pi}{12}\right)\,
\tanh\left(\frac{(\theta-\nu)}{2}+\frac{i\pi}{12}\right).
\end{equation}
After the substitution \eqref{classandquanparameter}, this expression is equivalent to the transmission factor ${^c}T$ for the classical defect problem linearized around the constant solution $u=v=\lambda=0$:
\begin{equation}
{^c}T=\frac{i(e^{\theta-\eta}-e^{-\theta+\eta})+
\sqrt{3}}{i(e^{\theta-\eta}-e^{-\theta+\eta})-\sqrt{3}}
=\tanh\left(\frac{(\theta-\eta)}{2}-\frac{i\pi}{3}\right)\,
\tanh\left(\frac{(\theta-\eta)}{2}-\frac{i\pi}{6}\right).
\end{equation}

\section{The $a_2^{(2)}$ transmission matrix: the linear equation}
\label{thelinearequation}

\p In this section, the transmission matrix for the type-II defect in the Tzitz\'eica model will be derived using a different technique based on an infinite-dimensional representation of the quantum algebra  underpinning the model. In fact, the $S$ matrix of the model is completely characterized by the algebra
$U_q(a_2^{(2)})$ \cite{smirnov, efthimiou} and  the non-local conserved charges that generate the  algebra have been constructed some time ago \cite{efthimiou}.

\p Choosing $\alpha_0$ to be the shorter root, the
generalized Cartan matrix of $a_2^{(2)}$ is taken to be
\begin{equation}\label{CartanMa22}
C_{ij}=\left(
  \begin{array}{cc}
    \phantom{*}2 & -1 \\
    -4 & \phantom{*}2 \\
  \end{array}
\right),\quad i,j=0,1.
\end{equation}
The quantized universal enveloping algebra associated with  $a_2^{(2)}$,
namely $U_q(a_2^{(2)})$ (see for instance \cite{jimbo94}), has six generators $\{X^{\pm}_1,X^{\pm}_0,K_1,K_0\}$.
The integers $d_i$, which symmetrize the matrix \eqref{CartanMa22} in the sense that $d_i\,C_{ij}=d_j\,C_{ji}$, are chosen to be $d_0=4$, $d_1=1$. Then,
the generators of the algebra satisfy
\begin{eqnarray}\label{uqa22comrel}
[K_1,K_0]=0,\quad K_i\,K_i^{-1}=K_i^{-1}\,K_i=1,\quad  i=0,1,\phantom{mmml}\nonumber\\
K_0\,X_0^{\pm}\,K_0^{-1}=q^{\pm 4}\,X_0^{\pm},\quad
[X_0^+,X_0^-]=\frac{K_0^2-K_0^{-2}}{q^4-q^{-4}}\phantom{mmmll}\nonumber\\
K_1\,X_1^{\pm}\,K_1^{-1}=q^{\pm 1}\,X_1^{\pm},\quad
 [X_1^+,X_1^-]=\frac{K_1^2-K_1^{-2}}{q-q^{-1}},\phantom{mmml}\nonumber\\
K_1X_0^{\pm}\,K_1^{-1}=q^{\mp 2}\,X_0^{\pm},\quad K_0\,X_1^{\pm}\,K_0^{-1}=q^{\mp 2}\,X_1^{\pm},
\quad [X_1^{\pm},X_0^{\mp}]=0,
\end{eqnarray}
together with the Serre relations,
\begin{eqnarray}\label{uqa22Serrerelations}
\nonumber&&\sum_{k=0}^5 (-1)^k
\begin{bmatrix}
5 \\
k
\end{bmatrix}
_q
\left(X_1^{\pm}\right)^{5-k} X_0^{\pm} \left(X_1^{\pm}\right)^{k}=0\\
&&\sum_{k=0}^2 (-1)^k
\begin{bmatrix}
2 \\
k
\end{bmatrix}
_{q^4}
\left(X_0^{\pm}\right)^{2-k} X_1^{\pm} \left(X_0^{\pm}\right)^{k}=0,
\end{eqnarray}
where the generalized binomial coefficient is defined by,
$$
\begin{bmatrix}
n \\
k
\end{bmatrix}
_t
=\frac{[n]_t!}{[n-k]_t!\, [k]_t!}\, ,\quad [k]_t=\frac{t^k-t^{-k}}{t-t^{-1}}\, .$$
Finally, the coproducts $\Delta,\ \Delta^\prime$ are given by:
\begin{equation}\label{coproduct}
\Delta(K_i)=K_i\otimes K_i,\quad \Delta(X^{\pm}_i)
=X_i^{\pm}\otimes K_i^{-1}+K_i\otimes X_i^{\pm} \quad i=0,1.
\end{equation}
and
$$\Delta'(K_i)=\Delta(K_i),\quad \Delta'(X^{\pm}_i)
=K_i^{-1}\otimes X_i^{\pm} +X_i^{\pm} \otimes K_i  \quad i=0,1.$$

\p The fundamental representation of this algebra is three-dimensional and can be constructed using the  spin-1 representation of the algebra $U_q(sl_2)$ with generators
$X_1^{\pm}$ and $K_1$ given by
\begin{equation}
K_1=\left(
  \begin{array}{ccc}
    q & 0 & 0 \\
    0 & 1 & 0 \\
    0 & 0 & q^{-1} \\
  \end{array}
\right),\quad
X_1^+=\left(
  \begin{array}{ccc}
    0 & 1 & 0 \\
    0 & 0 & 1 \\
  0 & 0 & 0 \\
  \end{array}
\right)
,\quad
X_1^-=(q+q^{-1})
\left(
  \begin{array}{ccc}
    0 & 0 & 0 \\
    1 & 0 & 0 \\
    0 & 1 & 0 \\
  \end{array}
\right).
\end{equation}
It follows that the remaining generators in this representation of the algebra $U_q(a_2^{(2)})$ are:
\begin{equation}
K_0=\left(
  \begin{array}{ccc}
    q^{-2} & 0 & 0 \\
    0 & 1 & 0 \\
    0 & 0 & q^{2} \\
  \end{array}
\right),\quad
X_0^+=\left(
  \begin{array}{ccc}
    0 & 0 & 0 \\
    0 & 0 & 0 \\
  1 & 0 & 0 \\
  \end{array}
\right)
,\quad
X_0^-=
\left(
  \begin{array}{ccc}
    0 & 0 & 1 \\
    0 & 0 & 0 \\
    0 & 0 & 0 \\
  \end{array}
\right).
\end{equation}

\p To calculate the defect transmission matrix an infinite-dimensional representation of the Borel subalgebra generated by $\{X_1^+,\,X_0^+,\,K_1,\,K_0\}$ is also required  in addition to the three-dimensional representation given above. Such a representation can be constructed using a pair of annihilation and creation operators, together with a `number' operator, that act on an infinite-dimensional space as follows
$$a|j\rangle =F(j)\,|j-1\rangle ,\quad a^\dag|j\rangle =|j+1\rangle ,\quad N|j\rangle =j|j\rangle ,\quad j\in \mathbb{Z}.$$
It is not necessary to assume there is a ground state, meaning that $F(j)$ should vanish for some particular $j$, since, in the present context, the quantity $j$ represents a topological charge, which can be any integer, positive, negative, or zero. Also,
$$aa^\dagger=F(N+1),\quad a^\dagger a=F(N),\quad aG(N)=G(N+1)a, \quad a^\dagger G(N)=G(N-1)a^\dagger,$$
where $G(N)$ is any function of the number operator.
Using these, the generators of the Borel subalgebra are taken to be:
$$X_1^+=a^\dag,\quad X_0^+=a \,a,\quad K_1=\kappa_1\, q^{N},\quad K_0=\kappa_0\,q^{-2N},$$
where $\kappa_0$ and $\kappa_1$ are constants. This choice satisfies the relations \eqref{uqa22comrel} but the Serre relations \eqref{uqa22Serrerelations} require in addition,
\begin{eqnarray}\label{uqa22Frecurrencerelations}
\nonumber\sum_{k=0}^5 (-1)^k
\begin{bmatrix}
5 \\
k
\end{bmatrix}
_q
 F(N+k)F(N+k+1)=0\\
\sum_{k=0}^2 (-1)^k
\begin{bmatrix}
2 \\
k
\end{bmatrix}
_{q^4}
 F(N+2k)=0,\phantom{mm}
\end{eqnarray}
which in turn require
\begin{equation}
a^\dag\,a=F(N)=\left(b_1\,(-)^N+c_1\right)\,q^{-2 N}+\left(b_2\,(-)^N+c_2\right)\,q^{2 N},\quad b_1\,c_2=b_2\,c_1.
\end{equation}
Here, it will be convenient to make the following choices,
$$c_1\equiv f_1 \,q^{-1},\quad c_2\equiv f_2\, q,\quad b_1=b_2\equiv 0,$$
hence the number function is taker to be
\begin{equation}\label{numberfunction}
F(N)=f_1 \,q^{-2 N-1}+f_2\, q^{2 N+1}.
\end{equation}
This is a first example of a common phenomenon: typically, when representations of Borel subalgebras are constructed in this fashion, using generalized creation and annihilation operators, the associated number functions satisfy a pair of recurrence relations. Often, as in this case, one of the relations is linear but sometimes both recurrence relations are nonlinear. 

\p
The system will be studied in the homogeneous gradation \cite{efthimiou} where  the rapidity dependence is carried by the generators $X_0^{\pm}$, namely
\begin{equation}
E_1=X^{+}_1,\quad F_1=X^{-}_1,\quad E_0=x\,X^{+}_0,
\quad F_0=x^{-1}\,X^{-}_0.
\end{equation}
For a discussion on different gradations in connection with the quantum affine Toda models see \cite{delius94}.

\p The aim of this section is to find the transmission matrix $T'$ as an intertwiner of the infinite-dimensional representation with space $\mathcal{V}$ and the three-dimensional representation with space $V$:
$$T'(z/x):\mathcal{V}_{z}\otimes V_{x}\rightarrow \mathcal{V}_{z}\otimes V_{x}.$$
This is achieved by solving the linear intertwining condition
\begin{equation}\label{lineareqforT}
T'\,\Delta(b)=\Delta'(b)\,T',
\end{equation}
where $b$ is any generator of the  Borel subalgebra.

\p Before seeking a solution to this equation a few considerations are in order. The $R'$ matrix underlying the $U_q(a_2^{(2)})$ algebra is defined as the following intertwiner
$$R'(x_1/x_2):V_{x_1}\otimes V_{x_2}\rightarrow V_{x_1}\otimes V_{x_2},$$
where $V$ is  a representation space of the algebra, satisfying the
linear intertwining condition
\begin{equation}\label{linearequforR}
R'\,\Delta(a)=\Delta'(a)\,R',
\end{equation}
where $a$ is a generator of the algebra $U_q(a_2^{(2)})$. For the purposes of this article, the representation used is the three-dimensional one introduced above. Consequently the matrix $R'$, the solution of \eqref{linearequforR}, is:
\begin{equation}
R'(x,q)\equiv R(x,q^{-1}),\quad x={x_1}/{x_2},
\end{equation}
where $R$ is the matrix \eqref{a22Rmatrix}.
Then, by construction, the quadratic equation
$$R'(x_1/x_2)\,T'(z/x_2)\,T'(z/x_1)=T'(z/x_1)\,T'(z/x_2)\,R'(x_1/x_2)$$
holds. However, note that this equation acts on the space $W'=\mathcal{V}_{z}\,\otimes\, V_{x_1}\,\otimes \,V_{x_2}$, while the quadratic equation \eqref{STT} acts on the space $W=V_{x_1}\,\otimes \,V_{x_2}\,
\otimes \,\mathcal{V}_{z}$.

\p Another important difference between the convention adopted in this section and that used in section (\ref{thequadraticequation}) concerns the way in which the generators, and therefore the matrices $R'$ and $T'$, act on the `in'-states. In fact, in the process of solving the linear equation \eqref{lineareqforT} the following convention on the \lq in' and \lq out' states has been used (for instance using the generator $X^\pm_i$)
$$X^\pm_i|n\rangle _{\mbox{\small{in}}}=
|m\rangle _{\mbox{\small{out}}}.$$
Taking this into account, consider the fundamental representation of the $U_q(a_2^{(2)})$ algebra and the following weight vectors,
$$\lambda_1=\left(
              \begin{array}{ccc}
                1 \\
                0 \\
                0\\
              \end{array}
            \right),\quad
            \lambda_2=\left(
              \begin{array}{ccc}
                0 \\
                1 \\
                0\\
              \end{array}
            \right),\quad
            \lambda_3=\left(
              \begin{array}{ccc}
                0 \\
                0 \\
                1\\
              \end{array}
            \right),
$$
 corresponding to the soliton charges $+1$, $0$, $-1$, respectively, and $\lambda_1$ is the highest weight vector.
 Note, in both the fundamental representation of the algebra $U_q(a_2^{(2)})$ and the infinite-dimensional representation of the Borel subalgebra the generators $X_1^+$ act as raising operators while the $X_0^+$  are lowering operators.

\p After some calculation, the general solution of the linear equation \eqref{lineareqforT} is:
\begin{equation}\label{lineareqTa22}
T'=\left(
  \begin{array}{ccc}
    a'\,q^{-2N}+a''\,q^{2N} & k \,q^{N}\,a & v\,a\, a \\
    j \,q^{-N}\,a^\dag & b & i \,q^{-N}\,a \\
    w\, a^\dag \,a^\dag & l \,q^{N}\,a^\dag & c'\,q^{2N}+c''\,q^{-2N} \\
  \end{array}
\right),
\end{equation}
where
\begin{eqnarray}
a'=-\frac{\kappa_1^{-2}}{q^2\,(q^4-q^{-4})}\,\frac{x}{z},\qquad a''=\kappa_1^{-2}\,f_2^2\,q^2\,(1-q^2)\,(1-q^4),\nonumber\\
c'=-\frac{\kappa_1^{2}}{q^2\,(q^4-q^{-4})}\,\frac{x}{z},\qquad c''=\kappa_1^{2}\,f_1^2\,(1-q^{-2})\,(1-q^{-4}),\phantom{l}\nonumber\\
b=-\frac{1}{q^2\,(q^4-q^{-4})}\,\frac{x}{z}+f_1\,f_2\,
(q^{-2}-1)\,(1-q^4),\phantom{mml}\nonumber\\
k=\kappa_1^{-1}\,f_2\,q\,(1-q^{4}),\qquad i=\kappa_1\,f_1\,(q^{-4}-1),\phantom{mmmml}\nonumber\\
j=\frac{\kappa_1^{-1}\,q}{(1+q^4)}\,\frac{x}{z},\qquad
l=\frac{\kappa_1}{(1+q^4)}\,\frac{x}{z},\phantom{mmmmml}\nonumber\\
v= 1,\qquad w=\frac{(1-q^2)}{(1+q^4)}\,\frac{x}{z}.\phantom{mmmmmmml}
\end{eqnarray}
Also, equation \eqref{lineareqforT} forces
$$\kappa_0^2\,\kappa_1^4=1,$$
and $\kappa_0=\kappa_1^{-2}$ is the choice that has been made to obtain the solution \eqref{lineareqTa22}.

\p At this point it is possible to compare the $T$ matrix \eqref{Ta22solution} and the matrix $T'$ \eqref{lineareqTa22}. In fact, because of the different conventions adopted to describe the manner in which the two matrices act on the \lq in' states, it is necessary to consider the matrix ${T'}^T $. Noting,
$${R'}^T(x,q)=-R(x^{-1},q),$$
 it is expected that $T$ and $T^\prime$ are related by
\begin{equation}\label{TTpa22}
{T'}^T(x,q)=-T(x^{-1},q^{-1}),
\end{equation}
where the change from $q$ to $q^{-1}$ is related to the changed ordering in the choices of the spaces $W$ and $W'$, as explained previously. Taking all of this into account, the relationship \eqref{TTpa22} holds with a suitable identification of the four free parameters in each matrix, and a suitable choice of the free function $\mu(\alpha)$ that appears in \eqref{Ta22solution}, but it does not hold in general.

\p To understand the origin of the free function appearing in the solution \eqref{Ta22solution} from an algebraic point of view, consider a slightly different infinite-dimensional representation of the Borel subalgebra constructed as follows,
\begin{equation}\label{Bsubala22ext}
X_1^+=a^\dag\,\mu(N),\quad X_0^+=a \,a\,\lambda(N),\quad K_1=\kappa_1\, q^{N},\quad K_0=\kappa_0\,q^{-2N},
\end{equation}
with, as before,
\begin{equation}\label{aadadonspace}
a|j\rangle =F(j)|j-1\rangle ,\quad a^\dag|j\rangle =|j+1\rangle ,\quad N|j\rangle =j|j\rangle.
\end{equation}
This representation clearly satisfies the commutation relations \eqref{uqa22comrel} but the functions $F(N)$, $\mu(N)$ and $\lambda(N)$
are, at present, unspecified.
The solution of the linear equation \eqref{lineareqforT} is:
\begin{equation}\label{lineqTmatrixa22ext}
T'=
\left(
  \begin{array}{ccc}
    A(N) & K(N)\,a & V(N)\,a\,a \\
    J(N)\,a^\dag &  B(N) & I(N)\,a \\
    W(N)\,a^\dag\,a^\dag & L(N)\,a^\dag & C(N) \\
  \end{array}
\right),
\end{equation}
with
$$A(N)=a'\,q^{-2N}+a''\,q^{2N},\qquad C(N)=c'\,q^{-2N}+a'\,\kappa_1^4\,q^{2N},
\qquad B(N)=b,\qquad \kappa_0^2\,\kappa_1^4=1,$$
$$K(N)=\frac{z}{x}\,\frac{a'}{\kappa_0\,\kappa_1}\,\frac{(1-q^4)}{q^2}\,q^{-N}\,
\left(\frac{D(N+1)-q^4\,D(N+2)}{\mu(N)\,F(N+1)}\right),\quad
J(N)=a'\kappa_1\,\frac{(1-q^4)}{q}\,q^{-N}\,\mu(N),$$
$$I(N)=\frac{z}{x}\,\frac{a'\,\kappa_1}{\kappa_0}\,\frac{(1-q^4)}{q}\,q^{N}\,
\left(\frac{D(N+2)-q^4\,D(N+1)}{\mu(N)\,F(N+1)}\right),\quad
L(N)=a'\kappa_1^3\,\frac{(1-q^4)}{q^2}\,q^{N}\,\mu(N),$$
$$V(N)=\frac{z}{x}\,\frac{a'}{\kappa_0}\frac{(1-q^8)}{q^2}\,\frac{D(N+2)}
{\mu(N+1)\,\mu(N)\,F(N+2)\,F(N+1)},$$
$$W(N)=a'\kappa_1^2\frac{(1-q^2)\,(1-q^4)}{q^2}\,\mu(N-2)\,\mu(N-1),$$
where
\begin{equation}\label{Dfunctiona22}
D(N)= \lambda(N)\,\mu(N-1)\,\mu(N-2)\,F(N)\,F(N-1).
\end{equation}
In addition, the linear equation provides the following constraints
\begin{eqnarray}
(q^4+q^{-4})\,D(N)\,D(N-1)-D(N)\,D(N-2)-D(N+1)\,D(N-1)&=&0,\label{con1}\\
(q^{-2}+q^4)\,D(N+1)-q^2\,D(N)-D(N+2)=Y'(N),\label{con2}\\
(q^{-2}+q^4)\,D(N+1)-D(N)-q^2\,D(N+2)=Y''(N),\label{con3}
\end{eqnarray}
with
\begin{eqnarray*}
Y'(N)&=&\frac{x}{z}\,\frac{\kappa_0}{a'}\,\frac{1}{(1-q^4)}
\,(b-a'\,\kappa_1^2-\kappa_1^{-2}\,c'\,q^{-4N}),\\
Y''(N)&=&\frac{x}{z}\,\frac{1}{a'\,\kappa_0\,\kappa_1^2}\,\frac{1}{(1-q^4)}
\,(b\,\kappa_1^{-2}-a'-a''\,q^{4N}).
\end{eqnarray*}
Then, constraints \eqref{con2} and \eqref{con3} fix the constants $b$, $c'$, $a''$ appearing in the matrix \eqref{lineqTmatrixa22ext}
\begin{eqnarray*}
b&=&\frac{z}{x}\,\frac{a'}{\kappa_0}\,d_0\,(1-q^4)^2\,(q^{-2}-1)+a'\,\kappa_1^2,\\
c'&=&\frac{z}{x}\,\frac{a'\,\kappa_1^2}{\kappa_0}\,d_1\,(q^{-8}-1)\,(1-q^4)\,(1-q^{2}),\\
a''&=&\frac{z}{x}\,a'\,\kappa_1^2\,\kappa_0\,d_2\,(1-q^{8})\,(1-q^4)\,(1-q^{2}),
\end{eqnarray*}
and the form of the function $D(N)$ \eqref{Dfunctiona22}
\begin{equation}\label{a22Dfunction}
D(N)=d_0+d_1\,q^{-4N}+d_2\,q^{4N}.
\end{equation}
Finally,
the constraint \eqref{con1}
forces
$$d_0^2\,q^4=d_1\,d_2\,(1+q^4)^2.$$
So far, the Serre relations \eqref{uqa22Serrerelations} have not been considered but they are expected to restrict the forms of the functions $F(N)$, $\mu(N)$ and $\lambda(N)$ in the infinite-dimensional representation \eqref{Bsubala22ext}. Interestingly, however, the linear equation \eqref{lineareqforT} has already take them into account. In fact, the Serre relations can be written in terms of the function $D(N)$ defined by eq\eqref{Dfunctiona22},
\begin{eqnarray*}
(q^4+q^{-4})\,D(N)\,D(N-1)-D(N)\,D(N-2)-D(N+1)\,D(N-1)&=&0,\\
(q-q^{-1})(D(N)-D(N+5))+(q^5-q^{-5})(D(N+4)-D(N+1))\ \ \ &&\\
+(q^5-q^{-5})(q^2-q^{-2})(D(N+2)-D(N+3))&=&0.
\end{eqnarray*}
The first of these coincides with the constraint \eqref{con1} while the second is implied by the constraints \eqref{con2} and \eqref{con3}, which can be verified using the following fact,
$$D(N)-D(N+2)=\frac{x}{z}\,\frac{1}{a'\,\kappa_1^2}\,\frac{1}{(1-q^2)\,(1-q^4)}
\,(\kappa_0^{-1}\,a''\,q^{4N}-\kappa_0\,c'\,q^{-4N}).$$
Hence, the $T'$ matrix \eqref{lineqTmatrixa22ext} has four free parameters ($d_1$, $d_2$, $a''$, $z$) - apart from $x$, which is related to the rapidity of the soliton solution - and a free function $\mu(N)$.
At this stage, and in a similar way to that which was suggested for the matrix \eqref{lineareqTa22}, the full solution \eqref{Ta22solution} can be linked to the solution \eqref{lineqTmatrixa22ext}. Also, notice that if in \eqref{Bsubala22ext} $\mu(N)=\lambda(N)=1$, then the solution \eqref{lineqTmatrixa22ext} coincides with the solution \eqref{lineareqTa22}.

\p It is interesting to note that since the Serre relations provide constraints for the function $D(N)$, which is the product of three different functions, it is possible to choose one of them, namely the number function $F(N)$, as follows
$$F(N)= a^\dag\,a=\varepsilon_1\,+\varepsilon_2\,q^{-2N}.$$
Then
$$ a^\dag\,a- q^2\,a\,a^\dag=\varepsilon_1\,(1-q^2),$$
which is the defining relation of a $q$-oscillator algebra,
as originally proposed in \cite{Macfarlane89,Biedenharn89}.


\p In section (\ref{thequadraticequation}) it was suggested that the presence of an extra field $\lambda$, defined up to a `gauge' transformation, and confined at the defect location in the Lagrangian density for the type-II defect, has a counterpart in the quantum theory arising from the possibility of freely choosing a function in the transmission matrix. This is a significant distinction between type-I and type-II defects. In fact, consider the sine-Gordon model, which unlike the $a_2^{(2)}$ Toda model can support both kinds of defect. A similar  calculation to the one performed in this section for the $a_2^{(2)}$ case
can be also carried out for the sine-Gordon model. Start, for instance, with the following infinite-dimensional representation of the Borel subalgebra of the $U_q(a_1^{(1)})$ quantum group,
$$X_1^+=a^\dag\,\mu(N),\quad X_0^+=a \,\lambda(N),\quad K_1=\kappa_1\, q^{N},\quad K_0=\kappa_0\,q^{-N},$$
where $a^\dag$ and $a$ are a pair of raising and lowering operators acting on an infinite-dimensional space as in \eqref{aadadonspace}. Then, the linear equation \eqref{lineareqforT} (the fundamental two-dimensional representation of the $U_q(a_1^{(1)})$ algebra is also needed)
provides a solution that depends on the function $\mu(N)$ with
\begin{equation}\label{Dfunctiona11}
D(N)=\mu(N-1)\,\lambda(N)\,F(N)=f_0+f_1\,q^{2N}+f_2\,q^{-2N},
\end{equation}
$F(N)$ being the number function.
Unlike the function $D(N)$ for the Toda $a_2^{(2)}$ case, all three constants $f_0$, $f_1$, $f_2$ in \eqref{Dfunctiona11} are free and unrelated to each other. The type-II defect corresponds to choosing all three constants different from zero. On the other hand, the type-I defect, which is permitted in the sine-Gordon model, corresponds to setting $f_1=f_2=0$ \cite{weston}. This has the effect of eliminating the possibility of freely choosing the function $\mu(N)$ in the transmission matrix. On the contrary, in this case $\mu(N)$ is determined up to a similarity transformation. This observation fits with the Lagrangian prescription because  there is no additional field $\lambda$ localized at the defect in the
classical Lagrangian density for the type-I defect \cite{bczlandau}.

\p As a final remark, it is interesting to notice that for the Toda $a_2^{(2)}$ model it is not possible to find the soliton scattering matrix inside the soliton-defect transmission matrix, as was the case for the sine-Gordon model of a type-II defect \cite{cz2010}. In fact, for no choices of the parameters is it possible to find any finite dimensional representation embedded within the infinite dimensional representation proposed in this article. On the other hand, a truncation of the infinite-dimensional representation is possible `in one direction'. For example, consider for simplicity the number function \eqref{numberfunction}. Then, by choosing a suitable value for the constant ratio $f_1/f_2$ an upper or lower boundary value for the integer label $j$ can be arranged, leading to semi-infinite representations.

\section{The $T$ matrices for the $a_2^{(1)}$ Toda model: the $U_q(a_2^{(1)})$ algebra}
\label{a21algebra}

\p In this section, the transmission matrices for the affine Toda model related to the
$a_2^{(1)}$ roots will be obtained from the quantum algebra point of view and the already known results for the type-I defect \cite{cz2007} will be recovered. In addition, the transmission
matrix for the type-II defect related to the classical Lagrangian density \eqref{defectlagrangian}, with
the setting B as explained in section (\ref{classicalsection}), will be calculated for the first time.

\p The algebra $U_q(a_2^{(1)})$ has nine generators $\{X^{\pm}_i,K_i\}$ with $i=1,2,3$. The integers $d_i$, which symmetrize the generalized Cartan matrix (see section (\ref{thelinearequation})) are chosen to be $d_i=2$, $i=1,2,3$. Then, the generators obey the following relations (see for instance \cite{jimbo94})
$$
[K_i,K_j]=0,\quad
[X_i^{\pm},X_j^{\mp}]=0,\quad [X_i^+,X_i^-]=\frac{K_i^2-K_i^{-2}}{q^2-q^{-2}}\quad i,j=1,2,3,$$
$$K_i\,K_i^{-1}=K_i^{-1}\,K_i=1,\quad K_i\,X_i^{\pm}\,K_i^{-1}=q^{\pm 2}\,X_i^{\pm},\quad K_i\,X_j^{\pm}\,K_i^{-1}=q^{\mp}\,X_j^{\pm}\, ,$$
together with the following Serre relations
\begin{eqnarray}\label{a11Serrerelations}
[X_i^{\pm},X_j^{\pm}]=0 \qquad\mbox{if}\quad C_{ij}=0 \phantom{mmmmm}\nonumber \\
\sum_{k=0}^2 (-1)^k
\begin{bmatrix}
2 \\
k
\end{bmatrix}
_{q^2}
\left(X_i^{\pm}\right)^{2-k} X_j^{\pm} \left(X_i^{\pm}\right)^{k}=0\qquad \mbox{if}\quad C_{ij}=-1,
\end{eqnarray}
where $C_{ij}$ is the generalized Cartan matrix.
The coproduct $\Delta$ was defined in \eqref{coproduct}.

\p The first fundamental representation, or soliton representation, of such an algebra is given by the following three-dimensional generators
\begin{equation*}
K_1=\left(
  \begin{array}{ccc}
    q & 0 & 0 \\
    0 &  q^{-1} & 0 \\
    0 & 0 & 1 \\
  \end{array}
\right),\quad
K_2=\left(
  \begin{array}{ccc}
    1 & 0 & 0 \\
    0 & q & 0 \\
    0 & 0 &  q^{-1}  \\
  \end{array}
\right)
,\quad
K_3=\left(
  \begin{array}{ccc}
    q^{-1} & 0 & 0 \\
    0 & 1 & 0 \\
    0 & 0 & q \\
  \end{array}
\right),
\end{equation*}
\begin{equation}\label{a21finiterep}
X_1^+=(X_1^-)^T=\left(
  \begin{array}{ccc}
    0 & 1 & 0 \\
    0 & 0 & 0 \\
  0 & 0 & 0 \\
  \end{array}
\right),\
X_2^+=(X_2^-)^T=\left(
  \begin{array}{ccc}
    0 & 0 & 0 \\
    0 & 0 & 1 \\
  0 & 0 & 0 \\
  \end{array}
\right)
,\
X_3^+=(X_3^-)^T=\left(
  \begin{array}{ccc}
    0 & 0 & 0 \\
    0 & 0 & 0 \\
  1 & 0 & 0 \\
  \end{array}
\right).
\end{equation}
The system will be studied in the  spin gradation, that is
\begin{equation}\label{spingradetiona21}
E_i=x^{2/3}\,X^{+}_i,\quad F_i=x^{-2/3}\,X^{-}_i\qquad i=1,2,3.
\end{equation}
Note that in the three-dimensional space the weight vectors are
$$\lambda_1=\left(
              \begin{array}{ccc}
                1 \\
                0 \\
                0\\
              \end{array}
            \right),\quad
            \lambda_2=\left(
              \begin{array}{ccc}
                0 \\
                1 \\
                0\\
              \end{array}
            \right),\quad
            \lambda_3=\left(
              \begin{array}{ccc}
                0 \\
                0 \\
                1\\
              \end{array}
            \right),\quad
$$
hence, with respect to the representation \eqref{a21finiterep} $\lambda_1$ is the highest weight vector.
Within the representation, these vectors correspond to the topological charges
\begin{equation}\label{weightsa21}
l_1=\frac{1}{3}\,(2\,\alpha_1+\alpha_2),\quad
 l_2=-\frac{1}{3}\,(\alpha_1-\alpha_2), \quad
 l_3=-\frac{1}{3}\,(\alpha_1+2\,\alpha_2),
 \end{equation}
respectively, expressed in terms of the simple roots of the $a_2^{(1)}$ algebra.

\p The $R'$ matrix solution of the equation \eqref{linearequforR} with representation \eqref{a21finiterep}
and spin gradation \eqref{spingradetiona21} has been provided by Jimbo, for example in \cite{jimbo}, and is conveniently summarized:
\begin{equation}\label{Rmatrixa21}
R'=
\left(
  \begin{array}{ccccccccccc}
    a & 0 & 0 & | & 0 & 0 & 0 & | & 0 & 0 & 0 \\
    0 & b & 0 & | & c^- & 0 & 0 & | & 0 & 0 & 0 \\
    0 & 0 & b & | & 0 & 0 & 0 & | & c^+ & 0 & 0 \\
    - & - & - &|& - & - & -&| & - & - & -  \\
    0 & c^+ & 0 & | & b & 0 & 0 & | & 0 & 0 & 0 \\
    0 & 0 & 0 & | & 0 & a & 0 & | & 0 & 0 & 0 \\
    0 & 0 & 0 & | & 0 & 0 & b & | & 0 & c^- & 0 \\
    - & - & - &|& - & - & -&| & - & - & -  \\
    0 & 0 & c^- & | & 0 & 0 & 0 & | & b & 0 & 0 \\
    0 & 0 & 0 & | & 0 & 0 & c^+ & | & 0 & b & 0 \\
    0 & 0 & 0 & | & 0 & 0 & 0 & | & 0 & 0 & a \\
  \end{array}
\right)
\end{equation}
with
$$a=x\,q^2-\frac{1}{x\,q^2},\quad b=x-\frac{1}{x},\quad c^{\mp}=\left(q^2-\frac{1}{q^2}\right)\,\left(x\right)^{\mp 1/3},\quad x=\frac{x_1}{x_2},
\quad q=e~^{-4\pi^2 i/\beta^2}.$$
Note, the following notation has been adopted
\begin{equation}\label{a21notation}
R^{kl}_{ij}|\lambda_i\,\lambda_j\rangle _{\mbox{\small{in}}}=
|\lambda_k\,\lambda_l\rangle _{\mbox{\small{out}}}.
\end{equation}

\p For the purposes of the present article the following infinite-dimensional representation of the Borel subalgebra appears to be the most appropriate,
\begin{equation}\label{harmonicoscia21}
K_i=\kappa_i\,q^{N_i-N_{i+1}},\qquad X_i^+=a_i^\dag\,a_{i+1},\qquad i=1,2,3,
\end{equation}
where $a_i,\,a_i^\dag$ are three independent sets of annihilation and creation operators, each acting along a direction represented by one of the unit vectors $\{e_1,e_2,e_3\}$.
This choice of representation is not unique and alternatives can be found in previous literature. In the case of $a_2^{(1)}$ and its generalisation to $a_n^{(1)}$, other representations may be found in \cite{Hayashi, bazhanov2002, Boos2010}, for example. The representation using three operators, or the generalization to $a_n^{(1)}$ using $n+1$ operators, has the advantage of maintaining the cyclic symmetry of the extended Dynkin diagram, though there is the  disadvantage that $a_1^{(1)}$ appears to be an exception. In fact, as will be demonstrated in the Appendix, $a_1^{(1)}$ in this context is naturally the first member of the  $c_n^{(1)}$ series. Of course, it is also possible to represent the $a_1^{(1)}$ using two operators but the number functions are less constrained.


\p Proceeding slightly differently from the previous case, the oscillator operators of \eqref{harmonicoscia21} are chosen to act on  infinite-dimensional spaces as follows,
$$a_i|j_i\rangle =g_i(j_i)|j_i-1\rangle ,\quad a_i^\dag|j_i\rangle =f_i(j_i)|j_i+1\rangle ,\quad N_i|j_i\rangle =j_i|j_i\rangle ,
\quad j_i\in \mathbb{Q}$$
with
$$a_i\,a^\dag_{i}|j_i\rangle =F(j_i)|j_i\rangle =f_i(j_i)\,g_{i}(j_i+1)|j_i\rangle ,\qquad i=1,2,3.$$
A typical defect state will then be represented in the tensor product of these three spaces by $|j_1,j_2,j_3\rangle$.
The Serre relations require the number functions $F_i(N_i)$ to be
$$F_i(N_i)=c_i^+\,q^{2N_i}+c_i^-\,q^{-2N_i}.$$
It is convenient to set $c_i^+\equiv (f^+_i)^2$ and $c_i^-\equiv -(f^-_i)^2$, so that the number functions are
\begin{equation}\label{numberfunctiona21}
F_i(N_i)=(f^+_i)^2\,q^{2N_i}-(f^-_i)^2\,q^{-2N_i},
\end{equation}
and to choose auxiliary functions $f_i(N_i)$ and $g_{i}(N_i)$ as follows,
\begin{equation}\label{a21fandgfunctions}
f_i(N_i)=(f^+_i\,q^{N_i}+f^-_i\,q^{-N_i}),\quad
g_{i}(N_i)=(f^+_i\,q^{N_i-1}-f^-_i\,q^{-N_i+1}),\quad i=1,2,3.
\end{equation}
The reasons for this choice of parametrization will become apparent later.

\p
The general solutions $T'$ of the linear equation \eqref{lineareqforT} with the fundamental representation \eqref{a21finiterep} and the infinite-dimensional representation \eqref{harmonicoscia21} in the spin gradation \eqref{spingradetiona21} is:
\begin{equation}\label{Tmatrixa21}
T'=\left(
  \begin{array}{ccc}
   A(N_1,N_2,N_3) & k\,q^{-N_3}\,a_1\,a_2^\dag
    & v\,q^{N_2}\,a_1\,a_3^\dag \\
    j\,q^{N_3}\,a_2\,a_1^\dag & B(N_1,N_2,N_3)
    & i\,q^{-N_1}\,a_2\,a_3^\dag \\
    w\,q^{-N_2}\,a_3\,a_1^\dag
    & l\,q^{N_1}\,a_3\,a_2^\dag & C(N_1,N_2,N_3)\\
  \end{array}
\right),
\end{equation}
with
\begin{eqnarray}\label{a21Tmatrixdiage}
A(N_1,N_2,N_3)&=&(a'\,q^{-N_1+N_2+N_3}+a''\,q^{N_1-N_2-N_3}),\nonumber\\
B(N_1,N_2,N_3)&=&(b'\,q^{N_1-N_2+N_3}+b''\,q^{-N_1+N_2-N_3}),\nonumber\\
C(N_1,N_2,N_3)&=&(c'\,q^{N_1+N_2-N_3}+c''\,q^{-N_1-N_2+N_3}).
\end{eqnarray}
The constants appearing in \eqref{Tmatrixa21} and \eqref{a21Tmatrixdiage} are:
\begin{eqnarray*}
a'=\left(\frac{x}{z}\right)^{4/3}\,\frac{q\,\kappa_3}{(1-q^4)^2},\quad
a''=\left(\frac{z}{x}\right)^{2/3}\,\frac{(1-q^4)\,\kappa_3}{q}\,(f^+_1\,f^-_2\,f^-_3)^2 ,\quad
(\kappa_1\,\kappa_2\,\kappa_0)^2=1,\\
b'=\left(\frac{x}{z}\right)^{4/3}\,\frac{q\,\kappa_1}{(1-q^4)^2\,\kappa_2},\quad
b''=\left(\frac{z}{x}\right)^{2/3}\,\frac{(1-q^4)\,\kappa_1}{q\,\kappa_2}\,(f^+_2\,f^-_1\,f^-_3)^2, \phantom{mmm}\\
c'=\left(\frac{x}{z}\right)^{4/3}\,\frac{q}{(1-q^4)^2\,\kappa_3},\quad
c''=\left(\frac{z}{x}\right)^{2/3}\,\frac{(1-q^4)}{q\,\kappa_3}\,(f^+_3\,f^-_1\,f^-_2)^2, \phantom{mmmm}\\
w=-(f_2^-)^2,\quad k=-\frac{(f_3^-)^2}{\kappa_2},\quad i=-(f_1^-)^2\,\kappa_1,\phantom{mmmmmmmm}\\
v=\left(\frac{x}{z}\right)^{2/3}\,\frac{1}{(1-q^4)},\quad
j=\left(\frac{x}{z}\right)^{2/3}\,\frac{1}{(1-q^4)\,\kappa_2},\quad
l=\left(\frac{x}{z}\right)^{2/3}\,\frac{\kappa_1}{(1-q^4)}.\phantom{mmm}
\end{eqnarray*}
This result should be compared with the transmission matrix $T$ for the type-I defect found in \cite{cz2007}, which was obtained by solving explicitly the quadratic triangle equations. The previous solution is expected to coincide with the above solution \eqref{Tmatrixa21} for some restriction of the free parameters.

 \p To facilitate the comparison, allow the matrix \eqref{Tmatrixa21} to act on the infinite-dimensional space, noting that although there are three sets of annihilation and creation operators, each associated with  one of the basis vectors $e_i$, the states on which they act are constrained. The reason for this is the following. The topological charges that can be deposited on a defect belong to the weight lattice and if a weight is written in terms of the basis,
 $$j=j_1\, e_1+j_2\, e_2+j_3\, e_3$$
 then
 $$j_1 + j_2 + j_3=0.$$

\p The matrix $T'$ should  be compared with the matrix $T$ provided in eq(B.24) in \cite{cz2007}, which is explicitly
\begin{equation}\label{a21Tmatrixcubicequ}
T_\gamma^\beta=\left(
\begin{array}{ccc}
q^{\gamma\cdot l_1}\,\delta_{\gamma}^{\beta}
  &\varepsilon\,x^{-2/3}\,q^{-\gamma\cdot l_3}\,\delta_{\gamma}^{\beta-\alpha_1}
  &\varepsilon^2\,x^{-4/3}
  \, \delta_{\gamma}^{\beta+\alpha_0}\phantom{} \\
   \varepsilon^2\,x^{-4/3}
  \,\delta_{\gamma}^{\beta+\alpha_1}
  &q^{\gamma\cdot l_2}\,\delta_{\gamma}^{\beta}
  &\varepsilon\,x^{-2/3}\,q^{-\gamma\cdot l_1}\,\delta_{\gamma}^{\beta-\alpha_2}\\
  \varepsilon\,x^{-2/3}\,q^{-\gamma\cdot l_2}\,\delta_{\gamma}^{\beta-\alpha_0}
  &\varepsilon^2\,x^{-4/3}
  \,\delta_{\gamma}^{\beta+\alpha_2}
  & q^{\gamma\cdot l_3}
  \,\delta_{\gamma}^{\beta}\\
\end{array}\right),
\end{equation}
where $\varepsilon$ is the defect parameter in the quantum theory.
The vectors $l_i$ are the fundamental weights \eqref{weightsa21}, which are rewritten in terms of the vectors $\{e_i\}$ as follows
\begin{equation*}
l_{1}=\frac{1}{3}(2\,e_1-e_2-e_3),\quad l_{2}=\frac{1}{3}(2\,e_2-e_3-e_1),\quad
l_{3}=\frac{1}{3}(2\,e_3-e_1-e_2).
\end{equation*}
The vector labels $\beta$ and $\gamma$ are weights and, since the components $(\gamma_1,\gamma_2,\gamma_3)$ of $\gamma$ are constrained by $\gamma_1+\gamma_2+\gamma_3=0$, $\gamma\cdot l_i=\gamma_i$.

\p The $R$ used previously  in \cite{cz2007} is related to the $R'$ matrix \eqref{Rmatrixa21} as follows: $${R'}^T(q,x)=R(q^2,x).$$
Hence, as explained in section (\ref{thelinearequation}), it is expected
${T'}^T(q,x)=T(q^{-2},x)$.

\p Set $f^+_1=f^+_2=f^+_3=0$ in \eqref{Tmatrixa21}, multiply by $(1-q^4)^2\,(x/z)^{-4/3}$, set $\kappa_1=\kappa_2=\kappa_3=1$ and take the transpose. Then, use the following similarity transformation acting on the three-dimensional space
$$T'\rightarrow U\,T'\,U^{-1},\quad U=\mbox{diag}\,[1,(f_1^-\,f_2^-)^{1/3}\,(f_3^-)^{-2/3},
(f_2^-)^{2/3}\,(f_1^-\,f_3^-)^{-1/3}].$$
Finally, send $q$ to $q^{-1/2}$ to find that
the solution \eqref{Tmatrixa21} becomes
\begin{equation}\label{Tmatrixa21with0step1}
\hat{T'}^j_i=\left(
  \begin{array}{ccc}
    q^{j_1}\,\delta^i_j & (x/z)^{-2/3}\,\epsilon\,q^{-j_3}\,\delta_j^{i-\alpha_1}
    & (x/z)^{-4/3}\, \epsilon^2\,\delta^{i+\alpha_0}_j\\
    (x/z)^{-4/3}\,\epsilon^2\,\delta_j^{i+\alpha_1} &q^{j_2}\,\delta^i_j
    &(x/z)^{-2/3}\,\epsilon\,q^{-j_1}\,\delta^{i-\alpha_2}_j  \\
  (x/z)^{-2/3}\,\epsilon\,q^{-j_2}\,\delta^{i-\alpha_0}_j
    & (x/z)^{-4/3}\, \epsilon^2\,\delta^{i+\alpha_2}_j & q^{j_3}\,\delta^i_j\\
  \end{array}
\right),
\end{equation}
with $\epsilon= -(1-q^4)\,(f_1^-\,f_2^-\,f_3^-)^{2/3}$. Then, the matrices \eqref{a21Tmatrixcubicequ} and \eqref{Tmatrixa21with0step1} coincide with $\gamma_i\equiv j_i$ and $\varepsilon\equiv \epsilon\,z^{2/3}$.
Similarly, setting $f^-_1=f^-_2=f^-_3=0$ in \eqref{Tmatrixa21}, the matrix (B.36) in \cite{cz2007} is recovered.

\p Finally, the $T'$ matrix with all parameters different from zero corresponds to the transmission matrix for the type-II defect in the setting B, as explained in section (\ref{classicalsection}). Alternatively, the setting A for the type II-defect, and the solutions (5.9) and (B.35) in \cite{cz2007} but associated with the type I-defect, can be constructed by starting with the following finite and infinite representations of the $U_q(a^{(1)}_2)$ and Borel subalgebra, respectively:
\begin{equation*}
K_1=\left(
  \begin{array}{ccc}
    q^{-1} & 0 & 0 \\
    0 &  q & 0 \\
    0 & 0 & 1 \\
  \end{array}
\right),\quad
K_2=\left(
  \begin{array}{ccc}
    1 & 0 & 0 \\
    0 & q^{-1} & 0 \\
    0 & 0 &  q  \\
  \end{array}
\right)
,\quad
K_3=\left(
  \begin{array}{ccc}
    q & 0 & 0 \\
    0 & 1 & 0 \\
    0 & 0 & q^{-1} \\
  \end{array}
\right),
\end{equation*}
\begin{equation*}
X_1^+=(X_1^-)^T=\left(
  \begin{array}{ccc}
    0 & 0 & 0 \\
    1 & 0 & 0 \\
  0 & 0 & 0 \\
  \end{array}
\right),\
X_2^+=(X_2^-)^T=\left(
  \begin{array}{ccc}
    0 & 0 & 0 \\
    0 & 0 & 0 \\
  0 & 1 & 0 \\
  \end{array}
\right),\
X_3^+=(X_3^-)^T=\left(
  \begin{array}{ccc}
    0 & 0 & 1 \\
    0 & 0 & 0 \\
  0 & 0 & 0 \\
  \end{array}
\right)
\end{equation*}
and
\begin{equation*}
K_i=\kappa_i\,q^{-N_i+N_{i+1}},\qquad X_i^+=a_i\,a_{i+1}^\dag,\qquad i=1,2,3.
\end{equation*}
Note that, the spectral parameters for the two representations are introduced as follows
\begin{equation*}
E_i=x^{-2/3}\,X^{+}_i,\quad F_i=x^{2/3}\,X^{-}_i,\qquad i=1,2,3.
\end{equation*}
Clearly, the intertwiner \eqref{linearequforR} provides the same $R'$ matrix \eqref{Rmatrixa21}, while the intertwiner for the Borel subalgebras \eqref{lineareqforT} yields the required solutions for $T'$.

\p As was the case for the sine-Gordon model, the scattering matrix $R'$ is embedded inside the transmission matrix $T'$. To be able to see this explicitly, the matrix \eqref{Rmatrixa21} is rewritten as follows, taking into account that $j_1+j_2+j_3=0$,
\begin{equation}
{T'}^k_j=\left(
  \begin{array}{ccc}
    A_j\,\delta^k_{j} & K_j\,\delta^{k+\alpha_1}_{j}
    & V_j\,\delta^{k-\alpha_0}_{j}\\
    J_j\,\delta^{k-\alpha_1}_{j} & B_j\,\delta^{k}_{j}
    & I_j\,\delta^{k+\alpha_2}_{j} \\
   W_j\,\delta^{k+\alpha_0}_{j}
    & L_j\,\delta^{k-\alpha_2}_{j} & C_j\,\delta^{k}_{j} \\
  \end{array}
\right)
\end{equation}
with
\begin{eqnarray*}\label{fullTmatrixa21}
A_j&=&q\,\kappa_3\,q^{-2j_1}+
\left(\frac{z}{x}\right)^{2}\,\frac{(1-q^4)^3\,\kappa_3}{q}\,(f^+_1\,f^-_2\,f^-_3)^2 \,q^{2j_1},\\
B_j&=&\frac{q\,\kappa_1}{\kappa_2}\,q^{-2j_2}+
\left(\frac{z}{x}\right)^{2}\,\frac{(1-q^4)^3\,\kappa_1}{q\,\kappa_2}\,(f^+_2\,f^-_1\,f^-_3)^2 \,q^{2j_2},\\
C_j&=&\frac{q}{\kappa_3}\,q^{-2j_3}+
\left(\frac{z}{x}\right)^{2}\,\frac{(1-q^4)^3}{q\,\kappa_3}\,(f^+_3\,f^-_1\,f^-_2)^2 \,q^{2j_3},\\
W_j&=&-\left(\frac{z}{x}\right)^{4/3}\,(1-q^4)^2\,(f_2^-)^2\,q^{-j_2}
(f^+_1\,q^{j_1}+f^-_1\,q^{-j_1})\,(f^+_3\,q^{j_3-1}-f^-_3\,q^{-j_3+1}),\\ K_j&=&-\left(\frac{z}{x}\right)^{4/3}\,\frac{(1-q^4)^2}{\kappa_2}
\,(f_3^-)^2\,q^{-j_3}\,(f^+_2\,q^{j_2}+f^-_2\,q^{-j_2})\,(f^+_1\,q^{j_1-1}-f^-_1\,q^{-j_1+1}),\\
I_j&=&-\left(\frac{z}{x}\right)^{4/3}\,(1-q^4)^2\,\kappa_1\,(f_1^-)^2\,\,q^{-j_1}
(f^+_3\,q^{2j_3}+f^-_3\,q^{-j_3})\,(f^+_2\,q^{j_2-1}-f^-_2\,q^{-j_2+1}),\\
V_j&=&\left(\frac{z}{x}\right)^{2/3}\,(1-q^4)\,q^{j_2}\,(f^+_3\,q^{j_3}+f^-_3\,q^{-2j_3})
\,(f^+_1\,q^{j_1-1}-f^-_1\,q^{-j_1+1}),\\
J_j&=&\left(\frac{z}{x}\right)^{2/3}\,\frac{(1-q^4)}{\kappa_2}\,q^{j_3}
\,(f^+_1\,q^{j_1}+f^-_1\,q^{-j_1})\,\,(f^+_2\,q^{j_2-1}-f^-_2\,q^{-j_2+1}),\\
L_j&=&\left(\frac{z}{x}\right)^{2/3}\,(1-q^4)\,\kappa_1\,q^{j_1}\,
(f^+_2\,q^{j_2}+f^-_2\,q^{-j_2})\,(f^+_3\,q^{j_3-1}-f^-_3\,q^{-j_3+1}).
\end{eqnarray*}
For simplicity set $\kappa_i=1$, $i=1,2,3$  and set $z=x_1$ and $x=x_2$. If $R'$ is to be found inside $T'$ it is required that the three-dimensional soliton representation is contained within the infinite-dimensional representation in terms of annihilation and creation operators.

\p In order to achieve this consider the generators $X_i^+$ defined in \eqref{harmonicoscia21} and require the following relations to hold
$$X_1^+|l_1\rangle =X_1^+|l_3\rangle =0,\quad X_2^+|l_1\rangle =X_2^+|l_2\rangle =0,
\quad X_3^+|l_2\rangle =X_3^+|l_3\rangle =0,$$
and
$$X_1^+|l_2\rangle \propto|l_1\rangle ,\quad X_2^+|l_3\rangle \propto|l_2\rangle ,\quad
X_3^+|l_1\rangle \propto|l_3\rangle ,\quad$$
where $l_i$, $i=1,2,3$ are the fundamental weights \eqref{weightsa21}. These relations imply
 $f_i^-=f_i^+\,q^{-8/3}$ for $i=1,2,3$. In addition, examining the finite set of entries labelled by fundamental weights, the non-zero elements of the $T'$ matrix \eqref{fullTmatrixa21} are:
$$K_1^2=I_2^3=W_3^1\equiv \Lambda \,c^+,\quad J_2^1=L_3^2=V_1^3\equiv \Lambda \,c^-,$$
$$A_1^1=B_2^2=C_3^3\equiv \Lambda \,a,\quad A_2^2=A_3^3=B_1^1=B_3^3=C_1^1=C_2^2\equiv \Lambda \,b,$$
where $c^+,\,c^-,\,a,\,b$ are the non zero elements of the $R'$ matrix \eqref{Rmatrixa21}. The
overall factor $\Lambda$ is
$$\Lambda=-\left(\frac{x_1}{x_2}\right)\,q^{5/3},\quad\mbox{with}\quad
f^+_1=f^+_2=f^+_0\equiv f^+,\quad (f^+)^2=-\frac{q^{14/3}}{(1-q^4)}.$$
Hence, the matrix $R'$ \eqref{Rmatrixa21} is recovered.

\p The infinite-dimensional representation can be also truncated in similar fashion to recover the three-dimensional anti-soliton representation corresponding to the other three-dimensional fundamental representation.
The weights of that representation are:
\begin{equation}\label{antiweightsa21}
h_{1}=-l_{3},\quad h_{2}=-l_{2},\quad h_{3}=-l_{1}
\end{equation}
where $h_{1}$ is the highest weight. Then
the following relations are required to hold
$$X_1^+|h_1\rangle =X_1^+|h_2\rangle =0,\quad X_2^+|h_1\rangle =X_2^+|h_3\rangle =0,
\quad X_3^+|h_2\rangle =X_3^+|h_3\rangle =0$$
and
$$X_1^+|h_3\rangle \propto|h_2\rangle ,\quad X_2^+|h_2\rangle \propto|h_1\rangle ,\quad
X_3^+|h_1\rangle \propto|h_3\rangle ,$$
implying $f_i^-=-f_i^+\,q^{2/3}$ with $i=1,2,3$. Taking into account only the vectors \eqref{antiweightsa21}, a calculation similar to the one performed above allows to establish all the non-zero elements of the scattering matrix between anti-solitons and solitons, namely
$$K_2^3=I_1^2=W_3^1\equiv \Lambda' \,(q^2-q^{-2})\,q^{-1}\,x^{1/3},
\quad J_3^2=L_2^1=V_1^3\equiv \Lambda' \,(q^2-q^{-2})\,q^{-1}\,x^{-1/3},\quad x=\frac{x_1}{x_2},$$
$$A_1^1=B_2^2=C_3^3\equiv -\Lambda' \,(x\,q^{3}-x^{-1}\,q^{-3}),\quad A_2^2=A_3^3=B_1^1=B_3^3=C_1^1=C_2^2\equiv -\Lambda' \,(x\,q-x^{-1}\,q^{-1}),$$
with
$$\Lambda'=\left(\frac{x_1}{x_2}\right)\,q^{10/3},\quad\mbox{and}\quad
f^+_1=f^+_2=f^+_0\equiv f^+,\quad (f^+)^2=-\frac{q^{4/3}}{(1-q^4)}.$$

\p It is worth emphasising that it is the possibility of being able to choose the functions $f_i$ and $g_i$ as in \eqref{a21fandgfunctions} that allows, with different choices of parameters, either of the finite fundamental representations to be found within the infinite-dimensional representation.

\section{The $T$ matrices for the $a_n^{(1)}$ Toda models: the $U_q(a_n^{(1)})$ algebra}
\label{an1T}

\p The calculations of section (\ref{a21algebra}) can be generalized to encompass the algebra $U_q(a_n^{(1)})$ for $n\geq 2$, in such a way as to obtain transmission matrices for type-I or type-II defects within the
$a_n^{(1)}$ affine Toda models. It is worth recalling that, in \cite{cz09}, the transmission matrices for type-I defects were written down guided by the type-I Lagrangian.

\p A fundamental $n+1$-dimensional representation for each algebra is:
\begin{eqnarray}\label{infinitedimrepan1}
K_i&=&\mbox{diag}\,[0,\dots,0,k_i,k_{i+1},0\dots,0],\quad k_i=q,\,\quad k_{i+1}=q^{-1},\nonumber\\
X^+_i&=&(X_i^-)^T=e_i\,\otimes \,e_{i+1},\qquad i=1,2\dots,h \ \ \hbox{modulo}(h),
\end{eqnarray}
where $h=n+1$ is the Coxeter number of the algebra.
These systems will be studied in the spin-gradation, which reads
\begin{equation}\label{evalra21}
E_i=x^{2/h}\,X^{+}_i,\quad F_i=x^{-2/h}\,X^{-}_i\qquad i=1,\dots,h.
\end{equation}

\p The infinite-dimensional representation of the Borel subalgebra, which generalizes the representation \eqref{harmonicoscia21}, is obtained by using $h$ annihilation and creation operators acting along the $h$ orthogonal directions represented by the unit vectors $\{e_1,e_2,\dots,e_{h}\}$. In terms of these, the generators are defined as follows
\begin{equation}\label{harmonicoscia2n}
K_i=\kappa_i\,q^{N_i-N_{i+1}},\qquad X_i^+=a_i^\dag\,a_{i+1}\qquad i=1,2\dots,h,
\end{equation}
and the number functions are
\begin{equation}\label{numberfunctiona21}
F_i(N_i)=f_i(N_i)\,g_{i}(N_i+1),\qquad i=1,2\dots,h,
\end{equation}
with
$$f_i(N_i)=(f^+_i\,q^{N_i}+f^-_i\,q^{-N_i}),\quad
g_{i}(N_i)=(f^+_i\,q^{N_i-1}-f^-_i\,q^{-N_i+1}).$$
Before exhibiting the $T'$ matrices, solutions of the linear equation \eqref{lineareqforT} for the representations presented in this section, a few comments on the notations used are in order. The $T'$ matrix entries will be denoted by ${T'}_{k\,l}$ for  $k,\, l=1,\,\dots,\,h$, where the indices $k,\,l$ simply reflect the positions of the entries in the matrix (and all indices are to be understood modulo $h$). For simplicity, the parameters $\kappa_i$ are chosen to be unity. This choice is compatible with the constraints that these constants have to satisfy (but it is not the only such choice).

\p Finally, the generalisation of the solution \eqref{Tmatrixa21} for any affine Toda model in the $a_n^{(1)}$
series can be written down explicitly. The diagonal entries are:
\begin{equation}
T'_{kk}=q^{1-N_k+\sum_{i=1\neq k}^h\,N_i}+(-)^{h-1}\,
\left(\frac{z}{x}\right)^{2}\,{(1-q^4)^h}
\,q^{N_{kk}}\,(f_k^+)^2\,\prod_{i=1\neq k}^h\,(f_i^-)^2.
\end{equation}
with
$$N_{kk}={N_k-1-\sum_{i=1\neq k}^h\,N_i}.$$
The off-diagonal entries with $k>l$ are:
\begin{equation}
T'_{kl}=(-)^{h-1-|k-l|}\,\left(\left(\frac{z}{x}\right)^{2/h}(1-q^4)\right)^{h-|k-l|}
\,q^{N_{kl}}
\,\prod_{i=1}^{h-1-|k-l|}\,(f_{i+l}^-)^2
\,\,a_k\,a_l^\dag,
\end{equation}
where
$$N_{kl}={\sum_{i=1}^{|k-l|-1}\,N_{i+k}-\sum_{i=1}^{h-1-|k-l|}\,N_{i+l}},$$
and the off-diagonal entries with $k<l$ are:
\begin{equation}
T'_{kl}=(-)^{|k-l|-1}\,\left(\left(\frac{z}{x}\right)^{2/h}(1-q^4)\right)^{|k-l|}
\,q^{N_{kl}}
\,\prod_{i=1}^{|k-l|-1}\,(f_{i+l}^-)^2
\,\ a_k\,a_l^\dag,
\end{equation}
with
$$N_{kl}={\sum_{i=1}^{h-|k-l|}\,N_{i+k}-\sum_{i=1}^{|k-l|-1}\,N_{i+l}}.$$
As mentioned previously, if $f_i^+=0$, $\forall i$ or $f_i^-=0$, $\forall i$, these matrices refer to transmission matrices for  type-I defects,
 otherwise they refer to  type-II defects.

\p As noticed in the $U_q(a_2^{(1)})$ case, it is possible to truncate the infinite-dimensional representations \eqref{infinitedimrepan1} in order to obtain finite-dimensional representations and this is done by arranging suitable zeros in one or other of the functions $f_i(N_i)$ or $g_i(N_i)$. In particular,  setting $f_i^-=f_i^+\,q^{-2(h+1)/h}$,\  $\forall i$, the first fundamental representation for each $a_n^{(1)}$ Toda model is found, while setting $f_i^-=-f_i^+\,q^{2/h}$,\  $\forall i$, the $n^{th}$ fundamental representation is obtained (each of these representations being $h$-dimensional). For these reasons, the soliton-soliton scattering matrix, referring to solitons in the first fundamental representation, and the soliton-anti-soliton S-matrix, where the corresponding anti-solitons lie in the $n^{th}$  fundamental representation, can be found as truncations of the transmission matrix given above.

\p The next natural question to ask concerns the other fundamental representations and whether they can be obtained in similar fashion by truncating infinite-dimensional representations. To discuss this question it is useful to write the weights of the $k^{th}$ fundamental representation in the $n+1$-dimensional basis. Then the weight vectors have components $1-k/h$ repeated $k$ times and $-k/h$ repeated $h-k$ times in any permutation. The highest weight vector for the $k^{th}$ representation is
$$(1-k/h,\dots,1-k/h, -k/h,\dots,-k/h),$$
where there are $k$ entries of the first kind and $h-k$ of the second kind, and $k=1,\dots,n$. Then, the generators $a_i^\dagger a_{i+1},\ i=1,\dots h$ act on the associated states by adding 1 to the $i^{th}$ component and $-1$ to the $(i+1)^{th}$ component and multiplying the state by $f_i(j_i)g_{i+1}(j_{i+1})$. Clearly it will not be possible (except for $k=1,n$) to ensure the action of the operators is either zero or another weight within the set of weights belonging to a particular fundamental representation. At best, new weights are generated by repeated application of the generators and it might be possible to arrange for the augmented set of weights to be a full set of weights for another (non-fundamental) representation. This phenomenon can be demonstrated by an example.

\p Consider the $2^{nd}$ fundamental representation with highest weight state
$$|1-2/h, 1-2/h, -2/h,\dots, -2/h\rangle.$$
The only operator that maps this to another weight in the same set is $a_h^\dagger a_1$ since
$$a_h^\dagger a_1|1-2/h, 1-2/h, -2/h,\dots, -2/h\rangle\sim |-2/h, 1-2/h, -2/h,\dots,-2/h, 1-2/h\rangle,$$
while all the others will either map the state to zero (by arrangement) or map it outside the weight set. The maximum number of zeros that can be imposed is achieved by requiring $g_i(-2/h)=0,\ i=1,\dots h$, but then
$$a_1^\dagger a_2|1-2/h, 1-2/h, -2/h,\dots, -2/h\rangle =f_1(1-2/h)g_2(1-2/h)|2-2/h, -2/h,\dots,-2/h\rangle,$$
which is not zero. On the other hand, the state on the right hand side represents a weight that is exactly twice a weight in the $1^{st}$ fundamental representation. Examining the whole collection of weights reveals that the set of weights in the 2$^{nd}$ fundamental representation is enhanced by $n+1$ new weights, each being the double of a weight in the 1$^{st}$ fundamental representation. Next, note that the additional state above is clearly annihilated by all the operators except $a_h^\dagger a_1$ and the latter gives
$$a_h^\dagger a_1|2-2/h, -2/h,\dots,-2/h\rangle=f_h(-2/h)g_1(2-2/h)|1-2/h, -2/h,\dots,-2/h, 1-2/h\rangle,$$
where the state on the right hand side is another weight within the $2^{nd}$ fundamental representation. This behaviour is readily checked for all states to conclude the infinite representation can be truncated not to the $2^{nd}$ fundamental representation but rather to a representation augmented by the additional set of $h$ weights.  This representation has dimension $h(h+1)/2$ and highest weight $(2-2/h, -2/h,\dots,-2/h)$. Inspection reveals this to be the set of weights for a representation that would be described for the group in terms of a second rank symmetric tensor (or a  horizontal two-box Young Tableau). The behaviour represented by this example is quite typical and instead of obtaining the fundamental representations (represented by antisymmetric tensors), one obtains the corresponding symmetric tensor representations (or their conjugates). For example, in $a_3^{(1)}$, it is possible to truncate the infinite dimensional representation on the $\bf 4, \overline 4, 10, \overline{10}$, representations and, therefore only certain S-matrices among the fundamental representations can be recovered by truncating a transmission matrix. These are
$$S_{44},\quad S_{\bar 4\bar 4},\quad S_{4\bar 4},\quad S_{64},\quad S_{6\bar 4}.$$
In particular, $S_{66}$ is missing from the list.

\p More generally, one might ask the question which finite representations can be obtained by truncating the infinite representation. Clearly not all of them can be, since only one-dimensional weight spaces are possible. Thus, for example, in $a_2^{(1)}$, the ${\bf 1,\ 3,\ \overline{3},\ 6,\ \overline{6},\ 10,\ \overline{10},}$ etc. are attainable (but not the self-conjugate representations such as ${\bf 8,\ 27,}$ etc.).

\section{Discussion}

\p In this article, connections between certain integrable defects, which can be sustained within some affine Toda field theories, and particular infinite dimensional representations of the quantum algebras underpinning these models have been explored. The discovery \cite{cz2009} that, in addition to Toda models in the $a_n^{(1)}$ series, the classical $a_2^{(2)}$ Toda field theory could also support a defect, at least  provided an additional degree of freedom was added to the system, led inevitably to wondering  how such a defect would be described in the quantum context. The first part of the present article has been devoted to identifying a suitable quantum transmission matrix $T$ able to describe the scattering between a type-II defect and a soliton of the $a_2^{(2)}$ Toda field theory. The final result has been established following two different approaches. First, a general solution of the purely transmitting Yang-Baxter equation has been found by using the bulk $S$-matrix associated with the model, and standard techniques have been employed to further constrain the solution found and to prove consistency with the classical picture. Second, a suitable infinite dimensional representation of the Borel subalgebra of $U_q(a_2^{(2)})$  has been constructed `ad hoc' in terms of a pair of generalized creation and annihilation operators, and the intertwining condition between this infinite dimensional representation, which carries the topological charge of the defect, and the finite three dimensional representation taking care of the soliton space, has been solved. It has been shown that the two approaches lead to the same result, lending support to the choice of infinite dimensional representation employed. In addition, it has been noted that, unlike the situation with the transmission matrices related to a type-I defect, the $T$ matrices associated with a type-II defect have a significant amount of freedom represented not only by free parameters but also by a completely free function of the topological charge stored at the defect location, which cannot be removed by a similarity transformation. It is tempting to link the extra free function to the additional field appearing in the classical Lagrangian density, which is defined only up to a `gauge' transformation, and which characterizes the type-II defect. This freedom also enters in the infinite dimensional representations of the Borel subalgebra, as has been demonstrated in section (\ref{thelinearequation}).
Finally, unlike the sine-Gordon case it should be noted that the bulk scattering $S$ matrix is not embedded within the infinite dimensional $T$ matrix related to a type-II defect.

\p The second part of the article has been spent on the affine Toda models with defects related to the $a_n^{(1)}$ Lie algebras. Both a classical and a quantum descriptions of the type-I defect within these models were already available in \cite{bcz2004,cz2007,cz09}. However, no investigations were made so far on the type-II defect. The results presented fill this gap.  The classical Lagrangian setting supporting a type-II defect within the Toda models in the $a_n^{(1)}$ series has been briefly presented in section (\ref{classicalsection}), and an exhaustive analysis of the associated quantum problem from a representation point of view has been carried out. Suitable infinite dimensional representations for the Borel subalgebras of $U_q(a_n^{(1)})$  have been constructed in terms of pairs of creation and annihilation operators, and the most general solutions for the intertwining condition have been calculated. The infinite dimensional representations adopted seem to differ from those available in the literature. On the other hand, they appear to be the most appropriate representations for the purposes of this article since they lead to previously obtained results on the transmission matrices for the type-I defect, and to new transmission matrices associated to the type-II defect. Finally, the possibility of truncating infinite dimensional representations to obtain finite dimensional representations is discussed, and it has been noticed that not all fundamental representations can be obtained this way.

\p It is natural to ask how the analysis of integrable defects and infinite dimensional representations in the context of massive integrable field theories could be extended to the other Toda models. A suitable classical Lagrangian setting is still missing for most of these models, though some results are available and will be presented elsewhere. From a quantum point of view, it is crucial to construct suitable infinite dimensional representations of the related Borel subalgebras and in the appendix a particular way to build such representations using sets of pairs of creation and annihilation operators is proposed  for all Toda models. The beauty of the method lies in two simple rules. Using these rules representations can be constructed, almost uniquely,  starting from the Dynkin-Kac diagrams of the associated affine Lie algebras. The representations are described explicitly, though a full analysis, in the sense of discovering complete expressions for the transmission matrices, is missing. One goal is to link these representations to specific integrable defects possessing a classical description and to understand better the link between the transmission matrices, defects and classical B\"acklund transformations first noted in \cite{bczlandau}. It has already been mentioned that different infinite dimensional representations are available in the literature for some of the quantum groups investigated here. It would be interesting to explore whether they might have a role in the Toda models in association with some specific integrable structure. In this context the full complement of inequivalent representations remains to be discovered.

\vskip .5cm
\p {\large \bf Acknowledgements}\\

\p We are grateful for several illuminating conversations with Robert Weston.  We also wish to express our gratitude to the UK Engineering and Physical Sciences
Research Council for its support under grant reference EP/F026498/1. One of us (EC) wishes to thank Peter Goddard and the Institute for Advanced Study, Princeton for hospitality during the later stages of this work.

\appendix
\section{Infinite dimensional representations}
\label{appendixA}

\p In the main text an infinite dimensional representation of the Borel subalgebra of $U_q(a_n^{(1)})$ was provided in terms of $n+1$ generalised annihilation and creation operators. It turns out that similar representations can be developed for all affine algebras. Though it appears there are several ways to build them by using sets of pairs of annihilation and creation operators, the generalizations provided here seem to be the most relevant for the kinds of applications described in this article. They share the common feature that the numbers of pairs of annihilation and creation operators involved is always equal to the number of links in the Dynkin-Kac diagram and the generator corresponding to a specific root involves a total number of operators (annihilation or creation) equal to twice the corresponding mark. Within each pair the role of annihilation and creation operator $(a_i,a_i^\dagger)$ can be interchanged provided also the corresponding number operator is reversed in sign ($N_i\rightarrow -N_i)$. In certain cases, the particular examples being $b_n^{(1)},\ d_n^{(1)}, n>4,\   a_{2n-1}^{(2)},$ the squares of annihilation and creation operators could be replaced by single operators instead. In all cases, the number functions are constrained by Serre relations and, since a set of annihilation and creation operators is assigned to each link in the Dynkin-Kac diagram, each number function will be constrained by having to satisfy a pair of recurrence relations. Generally, the recurrence relations are of differing order and degree. In many cases, at least one of the relations is linear and can be solved straightforwardly; in those circumstances the second relation provides nonlinear constraints on the parameters of the solution to the linear relation. In other cases, neither of the recurrence relations is linear and determining the general solution is harder. Some information is given in the text concerning this matter and more is given below in the section on simply-laced algebras. A fuller description together with an investigation of which kinds of finite representations can be described by restricting the number functions (along the lines described for $a_n^{(1)}$ in section (6)) will be provided elsewhere.

\p For the defining relations of the quantum algebras see for instance \cite{jimbo94}; for the set of data corresponding to the extended Dynkin-Kac diagrams see \cite{kacbook}.

\subsection{Non-simply laced Lie algebras}

\subsubsection{$a_{2n}^{(2)}$, $n> 1$}

\p In this case, there are three different root lengths, $\alpha_0$ being the shortest, $\alpha_n$ the longest. The extended Cartan matrix is:
\begin{equation}
C_{ij}= \left(\begin{array}{rrrrrrrrr}
        2&-1&0&..&0&0\\
        -2&2&-1&..&..&0\\
        0&-1&2&-1&..&..\\
       .. &..& .. &..& .. &..&\\
        0& .. &..&-1&2&-1\\
        0&0&..&0&-2&2\\
       \end{array}\right)\quad i,j=0,\dots, n,\nonumber
\end{equation}
the marks can be taken to be $\{n_i\}=\{1,1,\dots, 1,1/2\}$, and $\{d_i\}=\{8,4,\dots,4,2\}$. Then, operators can be assigned as follows,
\begin{eqnarray}\nonumber
&& X^+_0=(a_0)^2,\quad X_i^+=a_{i-1}^\dagger a_{i},\ i=1,\dots, n-1,\quad X^+_n=a^\dagger_{n-1}\\
&& K_0=q^{-4N_0},\quad K_i=q^{2(N_{i-1}-N_{i})}, \ i=1,\dots, n-1,\quad K_n=q^{2N_{n-1}}.\nonumber
\end{eqnarray}
If the roots $\alpha_1,\dots, \alpha_{n-1}$ are omitted, the representation reduces to the $a_2^{(2)}$ representation presented earlier in section (\ref{thelinearequation}) (though there $d_0=4,\ d_1=1$).

\subsubsection{$b_{n}^{(1)}$, $n>2$}

\p In this case, there are two different root lengths and $\alpha_n$ is the short simple root. The extended Cartan matrix has the form
\begin{equation}
C_{ij}= \left(\begin{array}{rrrrrrrrr}
        2&0&-1&0&..&..&0&0\\
        0&2&-1&0&..&..&..&0\\
        -1&-1&2&-1&0&..&..&..\\
        0&0&-1&2&-1&..&..&..\\
        ..&..&..&..&..&..&..\\
        0&..&..&..&..&-1&2&-2\\
        0&0&..&..&..&0&-1&2\\
       \end{array}\right)\quad i,j=0,\dots,n.\nonumber
\end{equation}
The marks are taken to be $\{n_i\}=\{1,1,2\dots,2\}$ while $\{d_i\}=\{2,\dots,2,4\}$, and the operators may be assigned as follows
\begin{eqnarray}\nonumber
 X^+_0=(a_0)^2,\ X^+_1=(a_1)^2,\ X^+_2=a_0^\dagger a_1^\dagger a_2^2,\ X_i^+=(a_{i-1}^\dagger)^2 (a_{i})^2,\ i=2,\dots, n-1,\ X^+_n=(a^\dagger_{n-1})^4\ \\
 K_0=q^{-N_0},\ K_1=q^{-N_1}, \  K_2=q^{(N_0+N_1-N_2)/2}, \ K_i=q^{(N_{i-1}-N_{i})/2}, \ i=2,\dots, n-1,\ K_n=q^{N_{n-1}}.\nonumber
\end{eqnarray}

\subsubsection{$a_{2n-1}^{(2)}$, $n>2$}

\p In this case, there are two different root lengths and $\alpha_n$ is the long simple root. The extended Cartan matrix has the form
\begin{equation}
C_{ij}= \left(\begin{array}{rrrrrrrrr}
        2&0&-1&0&..&..&0&0\\
        0&2&-1&0&..&..&..&0\\
        -1&-1&2&-1&0&..&..&..\\
        0&0&-1&2&-1&..&..&..\\
        ..&..&..&..&..&..&..\\
        0&..&..&..&..&-1&2&-1\\
        0&0&..&..&..&0&-2&2\\
       \end{array}\right)\quad i,j=0,\dots,n.\nonumber
\end{equation}
The marks are taken to be $\{n_i\}=\{1,1,2\dots,2,1\}$ while $\{d_i\}=\{4,\dots,4,2\}$, and the operators may be assigned as follows
\begin{eqnarray}\nonumber
 X^+_0=(a_0)^2,\ X^+_1=(a_1)^2,\ X^+_2=a_0^\dagger a_1^\dagger a_2^2,\ X_i^+=(a_{i-1}^\dagger)^2 (a_{i})^2,\ i=2,\dots, n-1,\ X^+_n=(a^\dagger_{n-1})^2\ \\
 K_0=q^{-2N_0},\ K_1=q^{-2N_1}, \  K_2=q^{N_0+N_1-N_2}, \ K_i=q^{N_{i-1}-N_{i}}, \ i=2,\dots, n-1,\ K_n=q^{N_{n-1}}.\nonumber
\end{eqnarray}

\subsubsection{$c_n^{(1)}$}

\p In this case, there are two different root lengths and the longer roots are taken to be $\alpha_0$ and $\alpha_n$. The extended Cartan matrix has the form
\begin{equation}
C_{ij}= \left(\begin{array}{rrrrrrrrr}
        2&-2&0&.. &..&0&0\\
        -1&2&-1&..&..&..&0\\
        0&-1&2&-1&..&..&..\\
        ..&..&..&..&..&..&..\\
        0&..&..&..&-1&2&-1\\
        0&0&..&..&0&-2&2\\
       \end{array}\right)\quad i,j=0,\dots,n.\nonumber
\end{equation}
By analogy with $a_{2n}^{(2)}$, the marks are taken to be $\{n_i\}=\{1/2,1,\dots,1,1/2\}$ while $\{d_i\}=\{2,4,\dots,4,2\}$, and the operators may be assigned as follows
\begin{eqnarray}\nonumber
&& X^+_0=a_0,\quad X_i^+=a_{i-1}^\dagger a_{i},\ i=1,\dots, n-1,\quad X^+_n=a^\dagger_{n-1}\\
&& K_0=q^{-2N_0},\quad K_i=q^{2(N_{i-1}-N_{i})}, \ i=1,\dots, n-1,\quad K_n=q^{2N_{n-1}}.\nonumber
\end{eqnarray}
 It should now be clear that when all the short simple roots are omitted, the Cartan matrix collapses to the $a_1^{(1)}$ extended Cartan matrix and the operators remaining are $ X^+_0=a_0,\
 X^+_1=a_0^\dagger$. In this sense, the sine-Gordon model could be regarded as being the first member of the  $c_n^{(1)}$ series.

\subsubsection{$d_{n+1}^{(2)}$, $n>1$}

\p In this case, there are two different root lengths and the shorter roots are taken to be $\alpha_0$ and $\alpha_n$. The extended Cartan matrix has the form
\begin{equation}
C_{ij}= \left(\begin{array}{rrrrrrrrr}
        2&-1&0&.. &..&0&0\\
        -2&2&-1&..&..&..&0\\
        0&-1&2&-1&..&..&..\\
        ..&..&..&..&..&..&..\\
        0&..&..&..&-1&2&-2\\
        0&0&..&..&0&-1&2\\
       \end{array}\right)\quad i,j=0,\dots,n,\nonumber
\end{equation}
the marks are taken to be $\{n_i\}=\{1,\dots,1\}$, $\{d_i\}=\{4,2,\dots,2,4\}$, and the operators may be assigned as follows
\begin{eqnarray}\nonumber
&& X^+_0=(a_0)^2,\quad X_i^+=a_{i-1}^\dagger a_{i},\ i=1,\dots, n-1,\quad X^+_n=(a^\dagger_{n-1})^2\\
&& K_0=q^{-2N_0},\quad K_i=q^{N_{i-1}-N_{i}}, \ i=1,\dots, n-1,\quad K_n=q^{2N_{n-1}}.\nonumber
\end{eqnarray}

\subsubsection{$g_{2}^{(1)}$}

\p In this case there are two root lengths and $\alpha_0$ is a long root. The relevant data is
the Cartan matrix
\begin{equation}
C_{ij}= \left(\begin{array}{rrrrrrrrr}
        2&-1&0\\
       -1&2&-3\\
        0&-1&2\\
        \end{array}\right)\quad i,j=0,1,2,\nonumber
\end{equation}
with marks $\{n_i\}=\{1,2,3\}$, and $\{d_i\}=\{2,2,6\}$. Then, the operators may be assigned as follows
$$ X^+_0=(a_0)^2,\  X^+_1=a_0^\dagger (a_1)^3,\  X^+_2=(a_1^\dagger)^6,$$
$$ K_0=q^{-N_0},\  K_1=q^{(N_0-N_1)/2},\  K_2=q^{N_1}.$$

\subsubsection{$d_{4}^{(3)}$}

\p In this case there are two root lengths and $\alpha_0$ is a short root. The relevant data is
the Cartan matrix
\begin{equation}
C_{ij}= \left(\begin{array}{rrrrrrrrr}
        2&-1&0\\
       -1&2&-1\\
        0&-3&2\\

        \end{array}\right)\quad i,j=0,1,2,\nonumber
\end{equation}
with marks $\{n_i\}=\{1,2,1\}$, and $\{d_i\}=\{6,6,2\}$. Then, the operators may be assigned as follows
$$ X^+_0=(a_0)^2,\  X^+_1=a_0^\dagger (a_1)^3,\  X^+_2=(a_1^\dagger)^2,$$
$$ K_0=q^{-3N_0},\  K_1=q^{3(N_0-N_1)/2},\  K_2=q^{N_1}.$$

\subsubsection{$f_{4}^{(1)}$}

\p In this case there are two root lengths and $\alpha_0$ is a long root. The relevant data is
the Cartan matrix
\begin{equation}
C_{ij}= \left(\begin{array}{rrrrrrrrr}
        2&-1&0&0&0\\
       -1&2&-1&0&0\\
        0&-1&2&-1&0\\
        0&0&-2&2&-1\\
        0&0&0&-1&2\\
        \end{array}\right)\quad i,j=0,\dots,4,\nonumber
\end{equation}
the marks $\{n_i\}=\{1,2,3,4,2\}$, and $\{d_i\}=\{2,2,2,4,4\}$. Then, the operators may be assigned as follows
$$ X^+_0=(a_0)^2,\  X^+_1=a_0^\dagger (a_1)^3,\  X^+_2=(a_1^\dagger)^2 (a_2)^4,\  X^+_3=(a_2^\dagger)^6(a_3)^2,\  X^+_4=(a_3^\dagger)^4,$$
$$ K_0=q^{-N_0},\  K_1=q^{(N_0-N_1)/2},\  K_2=q^{(N_1-N_2)/3},\  K_3=q^{(N_2-N_3)/2},\  K_4=q^{N_3}.$$

\subsubsection{$e_{6}^{(2)}$}

\p In this case there are two root lengths and $\alpha_0$ is a short root. The relevant data is
the Cartan matrix
\begin{equation}
C_{ij}= \left(\begin{array}{rrrrrrrrr}
        2&-1&0&0&0\\
       -1&2&-2&0&0\\
        0&-1&2&-1&0\\
        0&0&-1&2&-1\\
        0&0&0&-1&2\\
        \end{array}\right)\quad i,j=0,\dots,4,\nonumber
\end{equation}
the marks $\{n_i\}=\{1,2,3,2,1\}$, and $\{d_i\}=\{4,4,4,2,2\}$. Then, the operators may be assigned as follows
$$ X^+_0=(a_0)^2,\  X^+_1=a_0^\dagger (a_1)^3,\  X^+_2=(a_1^\dagger)^4 (a_2)^2,\  X^+_3=(a_2^\dagger)^3 a_3,\  X^+_4=(a_3^\dagger)^2,$$
$$ K_0=q^{-N_0},\  K_1=q^{(N_0-N_1)/2},\  K_2=q^{2(N_1-N_2)/3},\  K_3=q^{(N_2-N_3)},\  K_4=q^{-2N_3}.$$

\subsection{Simply laced Lie algebras}

\p In all the next examples $d_i=2$.

\subsubsection{$d_{4}^{(1)}$}

\p In this case all roots have the same length and the root corresponding to the central spot on the Dynkin diagram is $\alpha_4$. The marks are $\{n_i\}=\{1,1,1,1,2\}$. If a representation is required to preserve the symmetry of the extended Dynkin-Kac diagram then a possible choice is:
\begin{eqnarray}\nonumber
 X^+_i=(a_i)^2,  &&i=0,1,2,3,\quad X^+_4=a_0^\dagger a_1^\dagger a_2^\dagger a_3^\dagger,\\
 K_i=q^{-N_i}, &&i=0,1,2,3,\quad K_4=q^{\sum_0^3 N_i/2}.\nonumber
\end{eqnarray}
It is not difficult to generalise this to $d_n^{(1)}$, $n>4$ (in a similar fashion to  the representation provided above for $b_n^{(1)}$).

\subsubsection{$e_{6}^{(1)}$}

 \p This case can be thought of as a slight generalisation of the previous example; all roots have the same length, the root corresponding to the central spot on the Dynkin-Kac diagram is $\alpha_2$, the Cartan matrix is
 \begin{equation}
 C_{ij}=\left(\begin{array}{rrrrrrrrr}
        2&-1&0&0&0&0&0\\
       -1&2&-1&0&0&0&0\\
        0&-1&2&-1&0&-1&0\\
        0&0&-1&2&-1&0&0\\
        0&0&0&-1&2&0&0\\
        0&0&-1&0&0&2&-1\\
        0&0&0&0&0&-1&2\\
         \end{array}\right)\quad i,j=0,\dots,6,\nonumber
\end{equation}
with marks $\{n_i\}=\{1,2,3,2,1,2,1\}$. Then, a possible assignment of creation and annihilation operators is,
\begin{eqnarray}\nonumber
X^+_0=(a_0)^2,\ X^+_4=(a_4)^2, \ X^+_6=(a_6)^2,\ X^+_2=(a_1^\dagger)^2 (a_3^\dagger)^2( a_5^\dagger)^2,\\\nonumber
X^+_1=a_0^\dagger (a_1)^3,\ X^+_3=a_4^\dagger (a_3)^3,\ X^+_5=a_6^\dagger (a_5)^3,\phantom{mmmm}\\ \nonumber
K_0=q^{-N_0},\ K_4=q^{-N_4},\ K_6=q^{-N_6},\ K_2=q^{(N_1+N_3+N_5)/3}\phantom{mm}\\\nonumber
K_1=q^{(N_0-N_1)/2},\ K_3=q^{(N_4-N_3)/2},\ K_5=q^{(N_6-N_5)/2}.\phantom{mm}
\end{eqnarray}

\subsubsection{$e_{7}^{(1)}$}

\p Again the Dynkin-Kac diagram has a symmetry and all roots have the same length. The central spot on the diagram corresponds to the root $\alpha_3$, the Cartan matrix is given by
 \begin{equation}
 C_{ij}=\left(\begin{array}{rrrrrrrrr}
        2&-1&0&0&0&0&0&0\\
       -1&2&-1&0&0&0&0&0\\
        0&-1&2&-1&0&0&0&0\\
        0&0&-1&2&-1&0&0&-1\\
        0&0&0&-1&2&-1&0&0\\
        0&0&0&0&-1&2&-1&0\\
        0&0&0&0&0&-1&2&0\\
        0&0&0&-1&0&0&0&2\\
         \end{array}\right)\quad i,j=0,\dots,7,\nonumber
\end{equation}
with marks $\{n_i\}=\{1,2,3,4,3,2,1,2\}$. A possible operator representation of the Borel subalgebra is:
\begin{eqnarray}\nonumber
X^+_0=(a_0)^2,\ X^+_6=(a_6)^2,\ X^+_7=(a_7)^4, \ X^+_3=(a_2^\dagger)^3(a_7^\dagger)^2 (a_4^\dagger)^3,\phantom{mmmm}\\\nonumber
X^+_1=a_0^\dagger(a_1)^3,\ X^+_5=a_6^\dagger(a_5)^3, \ X^+_2=(a_1^\dagger)^2(a_2)^4, \ X^+_4=(a_5^\dagger)^2(a_4)^4,\phantom{mmm}\\\nonumber
K_0=q^{-N_0},\ K_6=q^{-N_6},\ K_7=q^{-N_7/2},\  K_3=q^{(N_2+N_7+N_4)/4},\phantom{mmmmm}\\\nonumber
K_1=q^{(N_0-N_1)/2},\ K_5=q^{(N_6-N_5)/2},\ K_2=q^{(N_1-N_2)/3},\ K_4=q^{(N_5-N_4)/3}.\phantom{mm}
\end{eqnarray}

\subsubsection{$e_{8}^{(1)}$}

\p This case has no symmetry, adjacent roots along the longer legs are labelled $0-7$, where $\alpha_0$ is furthest from the root $\alpha_5$  corresponding to the junction and $\alpha_8$ is the other root adjacent to $\alpha_5$; with this labeling the marks are $\{n_i\}=\{1,2,3,4,5,6,4,2,3\}$. A possible operator representation of the Borel subalgebra is:
\begin{eqnarray}\nonumber
X^+_0=(a_0)^2,\ X^+_1=a_0^\dagger(a_1)^3,\ X^+_2=(a_1^\dagger)^2(a_2)^4, \ X^+_3=(a_2^\dagger)^3(a_3)^5, X^+_4=(a_3^\dagger)^4(a_4)^6\phantom{}\\\nonumber
X^+_8=(a_8)^6,\ X^+_7=(a_7)^4,\ X^+_6=(a_7^\dagger)^2(a_6)^6, \ X^+_5=(a_4^\dagger)^5(a_6^\dagger)^4(a_8^\dagger)^3, \phantom{mmm}\\\nonumber
K_0=q^{-N_0},\ K_1=q^{(N_0-N_1)/2},\ K_2=q^{(N_1-N_2)/3},\  K_3=q^{(N_2-N_3)/4},\ K_4=q^{(N_3-N_4)/5}\\\nonumber
K_8=q^{-N_8/3},\ K_7=q^{-N_7/2},\ K_6=q^{(N_7-N_6)/4},\ K_5=q^{(N_4+N_6+N_8)/3}.\phantom{mm}
\end{eqnarray}
As before, in all cases, the associated number functions are restricted by the Serre relations to satisfy a series of recurrence relations, though these need not be straightforward to solve. Since all the above representations have the property that a set of creation and annihilation operators is associated with a link of the diagram, it follows each set of operators $a_k^\dagger, a^{\phantom{\dagger}}_k$ is associated with a pair of marks ($r_k,s_k$) from the set $\{n_i\}$. Suppose also that
$$a_k^\dagger a_k^{\phantom{\dagger}} = F_k(N_k).$$
Then, for the simply-laced algebras above, $F_k(N_k)$ satisfies a pair of recurrence relations in the combinations
\begin{eqnarray}\nonumber
R_k(N_k)=F_k(N_k-r_k+1)F_k(N_k-r_k+2)\cdots F_k(N_k)\\\nonumber
 S_k(N_k)=F_k(N_k-s_k+1)F_k(N_k-s_k+2)\cdots F_k(N_k)
\end{eqnarray}
which, in turn, are given by the linear combinations
$$R_k(N_k)=\sum_\lambda\, c_\lambda\lambda^{N_k},\quad S_k(N_k)=\sum_\mu\, d_\mu \mu^{N_k},$$
where the allowed $\lambda, \mu$ satisfy
$$\lambda^{s_k}=q^2\ \hbox{or}\ q^{-2}, \quad \mu^{r_k}= q^2\ \hbox{or}\ q^{-2},$$
and $c_\lambda, d_\mu$ are constants. If one of the relations is linear in $F_k(N_k)$ then it is straightforward to determine the constraints arising from the second relation and thence to determine $F_k(N_k)$ . However, this is not always the case.

\end{document}